\documentclass[10pt,prb,twocolumn,aps,amsmath,amssymb,showpacs]{revtex4}

\usepackage{graphicx}
\usepackage{dcolumn}
\usepackage{bm}
\usepackage{wrapfig}
\usepackage{amsmath,amsfonts,amssymb,bm}
\usepackage[usenames]{color}
\usepackage[title]{appendix}

\begin{document}

\title{Theory of magnon-polaritons in quantum Ising materials}

\author{ R.D. McKenzie}
\email[Correspondence should be addressed to R.D.M, ]{ryanmck.van@gmail.com}
\affiliation{Unaffiliated} 
\author{M. Libersky}
\affiliation{Division of Physics, Math, and Astronomy, California Institute of Technology, Pasadena CA 91125, USA}
\author{D.M. Silevitch}
\affiliation{Division of Physics, Math, and Astronomy, California Institute of Technology, Pasadena CA 91125, USA}
\author{T.F. Rosenbaum}
\affiliation{Division of Physics, Math, and Astronomy, California Institute of Technology, Pasadena CA 91125, USA}

\begin{abstract}

We present a theory of magnon-polaritons in quantum Ising materials, and develop a formalism describing the coupling between light and matter as an Ising system is tuned through its quantum critical point. The theory is applied to Ising materials having multilevel single-site Hamiltonians, in which multiple magnon modes are present, such as the insulating Ising magnet LiHoF$_4$. We find that the magnon-photon coupling strengths may be tuned by the applied transverse field, with the coupling between the soft mode present in the quantum Ising material and a photonic resonator mode diverging at the quantum critical point of the material. A fixed system of spins will not exhibit the diamagnetic response expected when light is coupled to mobile spins or atoms. Without the diamagnetic response, one expects a divergent magnon-photon coupling strength to lead to a superradiant quantum phase transition. However, this neglects the effects of damping and decoherence present in any real system. We show that damping and decoherence may block the superradiant quantum phase transition, and lead to weak coupling between the soft magnon mode and the resonator mode. The results of the theory are applied to experimental data on the model system LiHoF$_4$ in a microwave resonator.

\end{abstract}

\maketitle

\section{Introduction}

Although light-matter interactions were central to the development of quantum field theory, it is only recently that the interactions between microwave photons and magnetic materials have been explored in detail. Indeed, it was in 2009 that Imamo$\breve{\text{g}}$lu pointed out that strong coupling is achieved between resonant cavity photons and a spin ensemble in a coupled spin-photon system \cite{Imamoglu}. A short time later, interactions between a nanomagnet and microwave photons in a spherical resonator were investigated by Soykal and Flatt{\'e} \cite{SoykalFlattePRL,SoykalFlattePRB}. Since 2010, developments in microwave resonator technology have pushed forward our ability to explore fundamental aspects of quantum physics \cite{LiberskySM}, and have led to the rapid development of the new field of quantum magnonics, and associated hybrid quantum technologies \cite{LQReview,HuRoadmap,BhoiKim}.

In this paper, we develop a general, finite temperature, quantum field theory that may be used to study light-matter interactions, including interactions between a quantum system and an oscillator bath environment \cite{Dicke, Hopfield, DeLiberato, FV, CaldeiraLeggett, LeggettSB}. The formalism is applicable to materials having strong Ising interactions between their constituent atoms, or spins, and materials with complicated, multilevel, single-site Hamiltonians, such as the quantum Ising magnet LiHoF$_4$ \cite{Bitko, Chakraborty, MckenzieStamp}, which undergoes a ferromagnetic to paramagnetic quantum phase transition in an applied transverse field.

We analyze the transverse-field Ising model (TFIM) in the presence of an applied ac magnetic field along the easy axis of the material
\begin{align}
\label{eq:TFIM}
\mathcal{H} &= \mathcal{H}_{TFIM} + B_z \cos{(\omega t)} \sum_i J_i^z
\\ 
\label{eq:TFIMAC}
\mathcal{H}_{TFIM} &= -\frac{1}{2} \sum_{i \neq j} V_{ij} J_i^z J_j^z - B_x \sum_i J_i^x.
\end{align}
This simple model of a quantum material in a microwave resonator can be quantized to obtain a quantum optics model in which the spins couple to an effective photon momentum operator ($p \sim i(a^{\dagger}-a)$)
\begin{align}
\label{eq:Hp}
\mathcal{H} = \mathcal{H}_{TFIM} - i \alpha (a^{\dagger}-a) \frac{1}{\sqrt{N}} \sum_i J_i^z.
\end{align}

The magnetic insulating crystal LiHoF$_4$ is often considered an archetypical quantum Ising material, albeit with a strong hyperfine interaction between each holmium spin and its nucleus, and with the dominant coupling between spins being long range dipolar interactions \cite{Bitko}. The results of our theory are applied to LiHoF$_4$, and they accommodate the low energy electronuclear modes present in the material. The coupling between light and matter depends on the atomic density of the matter. We note that the spin density of LiHoF$_4$ is more than three times that of YIG, which has been a primary focus of quantum magnonics. 

Coupled light and matter modes will hybridize, forming polariton modes. The theory of polaritons, named as such, stemmed from Hopfield's work \cite{Hopfield}, although earlier work on coupled light-matter modes is present in the literature \cite{Tolpygo, Huang}. The quantum optics model, given by equation (\ref{eq:Hp}) shares similarities with the Hopfield model \cite{Hopfield}, as well as the Dicke model \cite{Dicke}, and quantum environment models such as the Caldeira-Leggett and spin boson Hamiltonians \cite{CaldeiraLeggett, LeggettSB} (see Appendix \ref{ap:Models} for more details). A primary difference between the model given by equation (\ref{eq:Hp}) and the models introduced by Dicke and Hopfield is that in equation (\ref{eq:Hp}) we are considering a fixed system of spins, and no diamagnetic term is present in the Hamiltonian. A system comprised by mobile spins or atoms, as in the Dicke and Hopfield models, will exhibit a diamagnetic response. As discussed below, the diamagnetic term in light-matter Hamiltonians has important consequences, so equation (\ref{eq:Hp}) should be considered a distinct model.

The diamagnetic term has been the source of considerable controversy. In the absence of the diamagnetic term, as one increases the light matter coupling strength, a superradiant quantum phase transition is expected to occur, in which photons spontaneously appear in the ground state of the system \cite{HeppLieb, Wang}. The presence of the diamagnetic term forestalls this transition \cite{Rzazewski}. Furthermore, with the diamagnetic term present, it was shown by De Liberato that as one increases the light-matter coupling, the light and matter modes will in fact decouple \cite{DeLiberato}. The source of this light-matter decoupling is the diamagnetic response which localizes the photon modes away from the matter, and shifts the resonant frequency of the light mode. As the coupling strength is increased, one finds the polaritonic modes have a predominately light or matter character.

The spin-photon Hamiltonian given in equation (\ref{eq:Hp}) leads to an effective magnon-photon Hamiltonian in which the diamagnetic term is absent. As the quantum Ising material is tuned through its critical point, the spectral weight of the soft mode, and hence the magnon- photon coupling strength, diverges. As no diamagnetic term is present, this ought to lead to a superradiant quantum phase transition. However, we show that including the effects of dissipation and decoherence of the magnon modes leads to a very different outcome. When environmental degrees of freedom are taken into account, the resulting dissipation and decoherence couple to the divergence of the soft mode, providing a new means to prevent superradiance. We substantiate this theoretical prediction in experiments on a model quantum Ising magnet.

The remainder of this paper is structured as follows: To begin, in Section \ref{sec:ResPhys}, we provide a brief discussion of the magnon-polariton propagator and the resonator transmission function. This provides a primary connection between theoretical work and experimental results. The magnon-polariton theory is then developed in Section \ref{sec:MPtheory}. Starting with equations (\ref{eq:TFIM}) and (\ref{eq:TFIMAC}), we derive the magnon-polariton propagator for the coupled light-matter system, and an effective bosonic Hamiltonian describing the system. The calculation is lengthy, so we begin Section \ref{sec:MPtheory} with a detailed summary of the steps involved.
 
Having obtained the magnon-polariton propagator, we discuss its application to calculating mode energies and spectral weights in Section \ref{sec:DisRes}, first in the absence of damping and then with frequency-independent (ohmic) damping of the magnon modes. This concludes the theoretical portion of this paper. 

In Section \ref{sec:Expt}, we compare the theory with experimental data on LiHoF$_4$ in a microwave resonator \cite{LiberskySM, Kovacevic}. An ansatz is used to account for decoherence of the spins comprising the collective magnon modes. With dissipation and decoherence accounted for, we are able to make quantitative comparisons between results of this model and experimental measurements. 

\section{Resonator Physics}
\label{sec:ResPhys}

In a resonator experiment, one measures transmission of photons through the resonator, which is determined theoretically by the magnon-polariton propagator. Our quantum optics model is analyzed making use of the imaginary time ordered magnon-polariton propagator of the coupled system \cite{MahanBook}
\begin{align}
\label{eq:MPprop}
D_{mp}(\tau) = \bigr\langle T_{\tau} \bigr(a^{\dagger}(\tau)+a(\tau)\bigr) 
\bigr(a^{\dagger}+a\bigr) \bigr\rangle,
\end{align}
where $\langle T_{\tau}\ \cdots \rangle$ is an imaginary time ordered thermal average taken over the light and matter degrees of freedom. The results of this theory are applied to the quantum Ising magnet LiHoF$_4$ in a microwave resonator. 

In a two-port microwave resonator experiment, one may measure transmission of photons through the resonator. The resonator transmission function is given by \cite{PozarBook,WallsMilburn}
\begin{align}
S_{21} = \frac{x_2^{OUT}}{x_1^{IN}}\biggr|_{x_2^{IN}=0},
\end{align}
where $x_{1,2}^{IN/OUT}$ is a measure of the incoming and outgoing light at the resonator ports $1$ and $2$. The transmission function is the ratio of the outgoing photons at port $2$ to the incoming photons at port $1$ when no light is incident at port $2$. We assume the resonator transmission function is related to the magnon-polariton propagator by \cite{Harder}
\begin{align}
\label{eq:ResTrans}
|S_{21}(\omega)|^2 \propto \text{Im}[D_{mp}^{ret}(\omega)], 
\end{align}
where the proportionality constant depends on details of the resonator. The retarded magnon-polariton propagator $D_{mp}^{ret}(\omega)$, or photon response function, is defined by $D_{mp}^{ret}(\omega)=\beta D_{mp}(i\omega_n \rightarrow \omega+i0^+)$, with
\begin{align}
D_{mp}(i\omega_n) 
= \frac{1}{\beta}\int_0^{\beta} d\tau \ e^{i\omega_n \tau} D_{mp}(\tau),
\end{align}
where $\omega_n = 2\pi n/\beta$ are Bose-Matsubara frequencies \cite{MahanBook}. The imaginary component of $D_{mp}^{ret}(\omega)$ corresponds to the energy absorbed by the resonator photons. The transmission data varies over many orders of magnitude, and will be presented on a logarithmic scale
\begin{align}
10 \log{|S_{21}|^2} = 10 \log{\bigr(A \text{Im}[D_{mp}^{ret}]\bigr)}.
\end{align}
The proportionality constant $A$ can be adjusted so that the scale of the experimental data matches that of the theoretical results. In what follows we set $A=1$, leaving a more quantitative comparison of the experimental resonator transmission and the theoretical results as a subject for future work.

The magnon-polariton propagator is defined in terms of photon position operators, $x \sim a^{\dagger} + a$, whereas in equation (\ref{eq:Hp}) the spins couple to a photon momentum operator $p \sim i(a^{\dagger}-a)$. One can show that a canonical transformation that swaps the photon position and momentum operators leads to an equivalent formulation of the model in which the spins couple to an effective position operator \cite{Leggett84} (see Appendix \ref{ap:coupling})
\begin{align}
\label{eq:Hx}
\mathcal{H} = \mathcal{H}_{TFIM} - \alpha (a^{\dagger}+a) \frac{1}{\sqrt{N}} \sum_i J_i^z.
\end{align}
This canonical transformation facilitates the calculation of the magnon-polariton propagator for the interacting spin-photon system.

The cooperativity of a light-matter system is defined by $C \equiv 4g_m^2/(\Gamma_m \Gamma_r)$, where $\Gamma_r$ and $\Gamma_m$ are the linewidths (or dampings) of the light and matter modes, respectively \cite{Zhang}. In this expression, the coupling, $g_m^2=\alpha^2 A_m$, is between a magnon mode and a light mode,  where $\alpha$ is the spin-photon coupling given in equation (\ref{eq:Hx}), and $A_m$ is the spectral weight of the relevant magnon mode. This expression for the magnon-photon coupling is derived in Section \ref{sec:MPprop}. When the coupling strength exceeds the damping of the system ($C>1$), the modes are said to be strongly coupled, and there will be coherent energy oscillations between the matter and the light. Regardless of whether or not the modes are strongly coupled, the use of perturbation theory and the rotating wave approximation requires $\eta = g/\omega <<1$. If $\eta>0.1$ the system is said to be in the ultrastrong coupling regime, and if $\eta>1$ the system is in the deep strong coupling regime \cite{Kockum}. Somewhat confusingly, a system in the ultrastrong, or deep strong, coupling regime may be weakly coupled if $C<1$.

We have provided a brief description of the magnon-polariton propagator, the resonator transmission function, and a discussion of the cooperativity of a light matter system. We will make use of this material in the development of the magnon-polariton theory, and the comparison between the theory and experimental results for LiHoF$_4$ in a microwave resonator. In the next section, we provide a detailed derivation of the magnon-polariton propagator beginning with the basic model given by equations (\ref{eq:TFIM}) and (\ref{eq:TFIMAC}).

\section{Magnon-Polariton Theory}
\label{sec:MPtheory}

Our goal in this section is a detailed derivation of the magnon-polariton propagator, beginning with the basic spin model given by equations (\ref{eq:TFIM}) and (\ref{eq:TFIMAC}). Prior to delving into the calculation, we provide a brief summary of the required steps, and the terms which appear as the theory develops.

In Section \ref{sec:Hsp}, we quantize the longitudinal ac magnetic field present in our basic model, assuming a plane wave basis for the photons, and we divide the spin Hamiltonian into its mean field part, and interactions between fluctuations about the mean field. The photon part of the resulting spin-photon Hamiltonian contains a term describing the instantaneous Zeeman energy of the spins in the ac field. The spin-photon interaction is given by an effective photon momentum operator ($p \sim i(a^{\dagger}-a)$) coupled to fluctuations of the spins about their mean field. A canonical transformation is used to swap the photon momentum operator for a photon position operator in the interaction. A phenomenological filling factor is introduced to account for the coupling between spins and photons in a resonator where the plane wave assumption may break down.

In Section \ref{sec:DS}, we discuss the dynamic susceptibility of a quantum Ising system having a multilevel single site Hamiltonian. The dynamic susceptibility is discussed in both the mean field (MF) and the random phase approximations (RPA). To go beyond the RPA, phenomenological damping parameters are introduced to account for damping of the magnon modes due to interactions between magnetic fluctuations, phonons, or any other environmental degrees of freedom. The dynamic susceptibility is central to the calculation of the magnon-polariton propagator.

In Section \ref{sec:AFT}, we return to the spin-photon Hamiltonian derived in Section \ref{sec:Hsp}. An auxiliary field is introduced to account for the interactions between magnetic fluctuations in the spin component of the Hamiltonian. A shift in the auxiliary field allows a trace to be performed over the microscopic spin degrees of freedom, resulting in an effective field theory which describes photons coupled to collective spin excitations, or magnons, present in the quantum Ising material. An expression for the propagator of the free auxiliary field is developed. The shift in the auxiliary field leads to a diamagnetic term in the photon component of the Hamiltonian, $\mathcal{H}_{\gamma}^D = D (a^{\dagger}+a)^2$, which shifts the frequency of the resonator mode. Although this diamagnetic term is present in an intermediate stage of the development of the theory, we find that the free auxiliary field propagator contains a term which restores the photon frequency to its original value in the final expression for the magnon-polariton propagator, given in Section \ref{sec:MPprop}, so the diamagnetic response term arising from the shift in the auxiliary field plays no role in the final theory.

In Section \ref{sec:Hgamma}, we consider the photon component of the magnon-photon Hamiltonian and derive the free photon propagator. Finally, in Section \ref{sec:MPprop}, we consider the full magnon-photon Hamiltonian and derive the magnon-polariton propagator for the coupled light-matter system in terms of the dynamic susceptibility of the quantum Ising material. The spectral representation of the dynamic susceptibility is used to derive an equivalent bosonic Hamiltonian for the light-matter system. This completes the derivation of the magnon-polariton propagator.

\subsection{Spin-Photon Hamiltonian}
\label{sec:Hsp}

We consider the transverse field Ising model (TFIM) in a longitudinal ac field, $\mathcal{H} = \mathcal{H}_{\gamma} +  \mathcal{H}_{TFIM} + \mathcal{H}_{int}$, where $\mathcal{H}_{\gamma}$ is the photon Hamiltonian, the TFIM Hamiltonian is given in (\ref{eq:TFIM}), and the interaction between the spins and the magnetic field is
\begin{align}
\mathcal{H}_{int} = B_z \cos(\omega t) \sum_i J_i^z.
\end{align}

The TFIM may be treated in mean field (MF) theory, $\mathcal{H}_{TFIM} = \mathcal{H}_{MF} + \mathcal{H}_{fl}$, where the MF Hamiltonian is
\begin{align}
\label{eq:HMF}
\mathcal{H}_{MF} = E_{gs} -H_z \sum_i J_i^z - B_x \sum_i J_i^x,
\end{align}
with $H_z = V_0 \langle J^z \rangle_{MF}$, where the zero wavevector component of the interaction between spins is $V_0 = \sum_j V_{ij}$. The constant contribution to the ground state energy, $E_{gs}= V_0 \langle J^z \rangle_{MF}^2/2$, will be dropped from the subsequent analysis. The MF spin polarization $\langle J^z \rangle_{MF}$ is determined self consistently from the MF Hamiltonian \cite{SuzukiBook, DuttaBook}. The energy of the interactions between fluctuations in the longitudinal MF spin polarization are given by
\begin{align}
\mathcal{H}_{fl} = -\frac{1}{2} \sum_{i \neq j} V_{ij} \delta J_i^z \delta J_j^z,
\end{align}
where the fluctuation operator is defined by $\delta J_i^z = J_i^z - \langle J^z \rangle_{MF}$.

We consider a single electromagnetic field mode, in which case
\begin{align}
\mathcal{H}_{\gamma} = \hbar \omega_r \biggr(a^{\dagger} a + \frac{1}{2}\biggr).
\end{align} 
Assuming the magnetic field is generated by a plane wave, the quantized ac magnetic field in a volume $V_{res}$ may be written
\begin{align}
B_z \cos(\omega t) \rightarrow \widehat{B}_z = -i
\frac{g_L \mu_B}{c} \sqrt{\frac{\hbar \omega_r}{2 V_{res} \epsilon_0}} (a^{\dagger}-a), 
\end{align}
where, on the right-hand side, the time dependence is implicit in the photon operators and the amplitude of the field depends on the photon density. The Land{\'e} g-factor and Bohr magneton written explicitly in the quantized expression were previously included in the definition of $B_z$. We assume photons with a wavelength much larger than the sample size so that $e^{i \boldsymbol{q} \cdot \boldsymbol{r}} \approx 1$, with $\omega_r = qc$. 

Transforming the spin operators to momentum space
\begin{align}
J_{\bf k}^z = \frac{1}{\sqrt{N}} \sum_i e^{i{\bf k} \cdot {\bf r_i}} J_i^z,
\end{align}
we find that the interaction is $\mathcal{H}_{int} = - i \alpha (a^{\dagger}-a) \delta J_0^z$, with
\begin{align}
\label{eq:MomentumCoupling} 
\alpha = g_L \mu_B \sqrt{\frac{\mu_0 \hbar \omega_r N}{2 V_{res}}}.
\end{align}
The interaction is between spin fluctuations and an effective momentum operator, $p \sim i(a^{\dagger}-a)$. In Appendix \ref{ap:coupling} we show that a canonical transformation that swaps the photon position and momentum operators leads to an equivalent formulation of the problem in which
\begin{align}
\label{eq:Hintx}
\mathcal{H}_{int} = - \alpha (a^{\dagger}+a) \delta J_0^z.
\end{align}

We have dropped a term linear in the photon operators from the interaction, $\widehat{B}_z N \langle J^z \rangle_0 = - \alpha (a^{\dagger}+a) \sqrt{N} \langle J^z \rangle_0$. This is the (instantaneous) MF Zeeman energy of the spins in the longitudinal ac magnetic field. We will reintroduce this term as part of the photon Hamiltonian in Section \ref{sec:Hgamma}. In a system with $n$ atoms per unit cell, the total number of atoms is $N=nV_{sample}/V_{cell}$. The interaction strength may then be written
\begin{align}
\label{eq:BareCoupling}
\alpha = \eta \sqrt{2\pi} \sqrt{\hbar \omega_r}\sqrt{\rho J_D}
\quad \text{with} \quad J_D = \frac{\mu_0 (g_L \mu_B)^2}{4\pi},
\end{align}
where in our plane-wave approximation the filling factor is $\eta=\sqrt{V_{sample}/V_{res}}$, and the spin density is $\rho=n/V_{cell}$. In YIG we have $\rho = 4.22 \times 10^{27} m^{-3}$, whereas in LiHoF$_4$ the value is $\rho=1.39 \times 10^{28} m^{-3}$, which is about 3.3 times the value in YIG. The dipolar energy scale of the LiHoF$_4$ system is given by $\rho J_D=13.52mK$. For a discussion of the magnon-photon coupling strength in YIG, see references \cite{Zhang, Flower}.

Our result for the filling factor was based on a plane-wave assumption. In a realistic model of a microwave resonator \cite{SoykalFlattePRL, SoykalFlattePRB}, the plane-wave assumption may break down, and the filling factor will depend on details of the resonator. One may express the filling factor as \cite{Zhang, Flower}
\begin{align}
\eta = \sqrt{\frac{\bigr(\int_{V_{sample}}\textbf{B}(\textbf{r}) \cdot \widehat{z} \ d\textbf{r}\bigr)^2 }
{V_{sample} \int_{V_{res}}\bigr(\textbf{B}(\textbf{r})\bigr)^2\ d\textbf{r}}},
\end{align}
where $\textbf{B}(\textbf{r})$ is the magnitude of the ac resonator field. In this work, we will treat the filling factor as a phenomenological parameter. The results of our theory are applied to experimental data on LiHoF$_4$ in a loop gap microwave resonator \cite{Libersky}.

\subsection{Dynamic Susceptibility}
\label{sec:DS}

The dynamic susceptibility of a quantum Ising material is central to the development of the magnon-polariton theory. We will make frequent use of the dynamic susceptibility and its spectral decomposition. We proceed to review the dynamic susceptibility in both the mean field and the random phase approximations (MF and RPA). For a more detailed discussion of the dynamic susceptibility of magnetic materials, see \textit{Rare Earth Magnetism} by Jensen and MacKintosh \cite{JensenMackintosh}.
 
The MF Hamiltonian for each spin, and the matrix elements of the longitudinal spin operator, may be expressed in terms of eigenstates and energies of the single site MF Hamiltonian given by equation (\ref{eq:HMF})
\begin{align}
\label{eq:MF}
\mathcal{H}_{MF_i} = \sum_m E_m | m \rangle \langle m|
\quad \text{and} \quad 
c_{mn} = \langle m | J^z | n\rangle_{MF}, 
\end{align}
where $\{E_m\}$ are the single site energy levels of the system, and $\{|m\rangle\}$ are the associated eigenstates. We drop the constant shift in the ground state energy, $E_{gs}$, from subsequent analysis.

The modes of the spin system, and their associated spectral weights, follow from the connected imaginary time correlation function, or Green function, $g(\tau) = -\langle T_{\tau} \delta J^z(\tau) \delta J^z \rangle_{MF}$, where $T_{\tau}$ is the imaginary time ordering operator. In MF theory, transforming the Green function to Matsubara frequency space ($\omega_n = 2\pi n/\beta$), we may write the MF Green function as \cite{JensenMackintosh, Stinchcombe}
\begin{align}
g(i\omega_n) = \frac{1}{\beta} \int_0^{\beta} e^{i\omega_n \tau} g(\tau) d\tau  = 
\widetilde{g}(i\omega_n) - g_{el} \delta_{i\omega_n,0},
\end{align}
where in the final expression the Green function is divided into an inelastic component, and the quasi-elastic diffusive pole of the system. The longitudinal MF dynamic susceptibility and the Green function are related by $\chi_0(\omega) = - \beta g(i\omega_r \rightarrow \omega + i0^+) = \widetilde{\chi}_0(\omega) + \chi_{el}^0 \delta_{\omega,0}$. In terms of the MF energy levels and matrix elements of the longitudinal spin operator, one may write the dynamic susceptibility as 
\begin{align}
\label{eq:chi0}
\widetilde{\chi}_0(z) &= \sum_{n > m}|c_{mn}|^2 p_{mn} 
\frac{2 E_{nm}}{E_{nm}^2-z^2} 
\\ \nonumber
\beta \chi_{el}^0 &= \sum_m c_{mm}^2p_m - \biggr[\sum_m c_{mm}p_m \biggr]^2.
\end{align}
The $p_{mn} = p_m-p_n$ are differences between population factors $p_m = e^{-\beta E_m}/Z_{MF}$, where $Z_{MF} = \text{Tr}[e^{-\beta \mathcal{H}_{MF_i}}]$. The poles of $\widetilde{\chi}_0(z)$, $E_{nm} = E_n-E_m$, are the MF modes of the system, and their spectral weights are $a_{mn} = |c_{mn}|^2 p_{mn}$. The elastic contribution to the dynamic susceptibility, $\chi_{el}^0$, vanishes in the paramagnetic phase of the system ($c_{mm}=0$), and decays exponentially with temperature $\bigr(\chi_{el}^0 \sim T e^{(-E_1/T)} \bigr)$.

In the random phase approximation (RPA), the result for the dynamic susceptibility is $\chi(\boldsymbol{k},z) = \chi_0(z)/(1-V_{\boldsymbol{k}} \chi_0(z))$. One may solve for the poles of this function, and their residues, in order to obtain its spectral representation
\begin{align}
\label{eq:specrep}
\chi(\boldsymbol{k}, z) =  \sum_m  \biggr[\frac{A_{\boldsymbol{k}}^m 2E_{\boldsymbol{k}}^m}
{(E_{\boldsymbol{k}}^m)^2-z^2} \biggr] + \chi_{\boldsymbol{k}}^{el} \delta_{z,0},
\end{align}
where $A_{\boldsymbol{k}}^m$ is the spectral weight of the $m^{th}$ RPA mode $E_{\boldsymbol{k}}^m$. 

In the magnon-polariton theory, the wavelengths of the microwave photons are much larger than the size of the sample, so we are interested in the $\boldsymbol{k}=0$ limit of the dynamic susceptibility. In this limit we write $\chi(z) = \chi(\boldsymbol{k}=0, z)$, and we define $\{ \omega_m \} = \{ E_{\boldsymbol{k}=0}^m \}$, and $\{ A_m \} = \{ A_{\boldsymbol{k}=0}^m \}$, as the zero wavevector component of the magnon modes and their spectral weights. The spectral weights of the magnon modes are inversely proportional to the mode frequencies $A_m \sim 1/\omega_m$ (see Appendix \ref{ap:LiHo}), with the spectral weight of the soft mode diverging at the critical point of the system.

The RPA expression for the dynamic susceptibility neglects any damping of the magnon modes. In reality, the modes are damped by interactions between the magnetic fluctuations, and environmental degrees of freedom such as phonons, and extraneous photons inside a resonator. If the modes are assumed to behave as damped harmonic oscillators, the dynamic susceptibility may be written ($\chi_{el}=0$)
\begin{align}
\chi(\omega) =  \sum_m \frac{A_m 2\omega_m}{\omega_m^2-\omega^2 - i\omega\Gamma_m}. 
\end{align}
We have analytically continued to real frequencies $z\rightarrow \omega + i0^+$, and introduced the phenomenological damping parameters $\{ \Gamma_m \}$. As will be shown, the magnon-polariton propagator may be written in the same way. In terms of its reactive and absorptive parts ($\chi = \chi' + i\chi''$), the dynamic susceptibility is
\begin{align}
\label{eq:chi1}
\chi'(\omega) = \sum_m \frac{A_m 2\omega_m (\omega_m^2-\omega^2)}
{(\omega_m^2-\omega^2)^2 + (\omega \Gamma_m)^2} 
+ \frac{(\Gamma_0/2)^2 \chi_{el}}{\omega^2+(\Gamma_0/2)^2}
\end{align}
and
\begin{align}
\label{eq:chi2}
\chi''(\omega) = \sum_m \frac{A_m 2\omega_m \omega \Gamma_m}
{(\omega_m^2-\omega^2)^2 + (\omega \Gamma_m)^2} 
+ \frac{\omega \Gamma_0/2 \chi_{el}}{\omega^2+(\Gamma_0/2)^2},
\end{align}
where we have included the contribution from $\chi_{el}$ to illustrate its role in the theory. 

The damping parameter will downshift the resonant frequency of the mode, $\omega_m \rightarrow \widetilde{\omega}_m = \sqrt{\omega_m^2-(\Gamma_m/2)^2}$, and if the damping exceeds the mode energy, $\Gamma_m/2 > \omega_m$, the mode becomes overdamped. The shift in the mode energy may be eliminated by introducing a counterterm to the theory. This is accomplished by setting $z=\omega+i\Gamma_m/2$ for each mode in equation (\ref{eq:specrep}). The dynamic susceptibility is then
\begin{widetext}
\begin{align}
\label{eq:chiprime}
\chi'(\omega) = \sum_m  \biggr[\frac{A_m (\omega+\omega_m)}{(\omega+\omega_m)^2 + (\Gamma_m/2)^2}
-\frac{A_m (\omega-\omega_m)}{(\omega-\omega_m)^2 + (\Gamma_m/2)^2}\biggr]
+\frac{\chi_{el} (\Gamma_0/2)^2}{\omega^2+(\Gamma_0/2)^2} , 
\end{align}
and
\begin{align}
\label{eq:chiprime2}
\chi''(\omega) = \sum_m 
\biggr[\frac{A_m \Gamma_m/2}
{(\omega-\omega_m)^2 + (\Gamma_m/2)^2}
-\frac{A_m \Gamma_m/2}{(\omega+\omega_m)^2 + (\Gamma_m/2)^2}\biggr]
+\frac{\chi_{el} \omega \Gamma_0/2}{\omega^2+(\Gamma_0/2)^2} . 
\end{align}
\end{widetext}
With the counterterm present, the effect of the damping is to broaden the delta function peaks associated with absorption and emission by the magnon modes into Lorentzians. Damping also broadens the quasielastic diffusive pole into an additional peak in the absorption spectrum, albeit with a different lineshape. The elastic contribution to the dynamic susceptibility vanishes in the paramagnetic phase of the system, and decays exponentially with temperature. In the time domain, the Lorentzian function describes exponentially decaying oscillations at a fixed frequency, rather than the strictly exponential decay of excitations seen in an overdamped harmonic oscillator.

We make use of the spectral representation of the dynamic susceptibility to calculate the magnon-photon coupling strengths in the magnon-polariton theory.

\subsection{Auxiliary Field Theory}
\label{sec:AFT}

To derive the magnon-photon Hamiltonian, we make use of the partition function as a means to renormalize the system. The interactions between spins are decoupled via the introduction of an auxiliary Hubbard-Stratonovich field, which allows us to average out the microscopic spin degrees of freedom. The resulting theory describes photons coupled to the collective spin excitations, or magnons, present in the material.

We divide the total Hamiltonian of the spin-photon system into two terms $\mathcal{H} = \mathcal{H}_0 + \mathcal{H}'$, where $\mathcal{H}'$ contains the spin fluctuations
\begin{align}
\mathcal{H}' = -\frac{1}{2} \sum_{i \neq j} V_{ij} \delta J_i^z \delta J_j^z 
- \alpha (a^{\dagger}+a) \frac{1}{\sqrt{N}} \sum_i \delta J_i^z
\end{align}
and $\mathcal{H}_0 = \mathcal{H}_{MF} + \mathcal{H}_{\gamma}$. The photon Hamiltonian contains a contribution from the instantaneous Zeeman energy of the spins in the ac field as discussed following equation (\ref{eq:Hintx}). 

The partition function, written in the Matsubara formalism, is given by \cite{MahanBook}
\begin{equation}
\label{eq:PF}
Z = Z_{\mathcal{H}_0} \biggr\langle T_{\tau} \exp 
\biggr[-\int_{\tau} \beta \mathcal{H}'(\tau) \biggr] \biggr\rangle_{0},
\end{equation}
where $\int_{\tau} \equiv \int_0^{\beta} d\tau / \beta$. The interactions between spin fluctuations may be decoupled via the introduction of an auxiliary Hubbard-Stratonovich field \cite{MckenzieStamp}
\begin{align}
\frac{Z}{Z_{\mathcal{H}_0}} = \int \mathcal{D}\phi 
\exp\biggr(&-\frac{1}{2}\int_{\tau}  \sum_{\boldsymbol{k}} 
|\phi_{\boldsymbol{k}}(\tau)|^{2}\biggr) 
\\ \nonumber
& \times \biggr\langle T_{\tau} \exp\biggr(\int_{\tau} V(\tau)\biggr) \biggr\rangle_0,
\end{align}
where the integration measure is $\mathcal{D\phi} = d\phi_{\boldsymbol{k}}/\sqrt{2\pi}$, and (suppressing the $\tau$ dependence)
\begin{align}
V = \sum_{\boldsymbol{k}} \biggr[\phi_{-\boldsymbol{k}} \sqrt{\beta V_{\boldsymbol{k}}} 
+ \beta \alpha [a^{\dagger}+a]\delta_{\boldsymbol{k},0}\biggr]  \delta J_{\boldsymbol{k}}^z.
\end{align}

We proceed by shifting the auxiliary field so that the dependence of the interaction on the photons is in the Gaussian prefactor
\begin{align}
\label{eq:fieldshift}
\phi_0 \rightarrow \phi_0 - \frac{\beta \alpha (a^{\dagger}+a)}{\sqrt{\beta V_0}}.
\end{align}
Multiplying out the result for the zero wavevector component of the Gaussian prefactor, the partition function is 
\begin{align}
\frac{Z}{Z_{\mathcal{H}_0}} &= \biggr\langle \biggr\langle T_{\tau} 
\int \mathcal{D}\phi 
\exp\biggr( \int_{\tau} \alpha_{\phi} \phi_0 (a^{\dagger}+a) \biggr)
\\ \nonumber &\times
\exp\biggr(-\frac{1}{2}\int_{\tau} \sum_k |\phi_{\boldsymbol{k}}|^{2}\biggr)
\times \exp\biggr(\int_\tau V_s \biggr) \biggr\rangle_s\biggr\rangle_{\gamma},
\end{align}
where the dimensionless coupling between the photon operators and the magnetic fluctuations is $\alpha_{\phi} = \beta \alpha / \sqrt{\beta V_0}$. The interaction between the shifted auxiliary field and the spin fluctuations is
\begin{align}
V_s(\tau) = \sum_k \phi_{-\boldsymbol{k}}(\tau) \sqrt{\beta V_{\boldsymbol{k}}}\ 
\delta J_{\boldsymbol{k}}^z(\tau).
\end{align}
The thermal average over the eigenstates of $\mathcal{H}_0$ has been written in terms of separate averages over the spin and photon eigenstates, $\langle \cdots \rangle_0 = \langle\langle \cdots \rangle_s \rangle_{\gamma}$. This is possible because in $\mathcal{H}_0$ the Hilbert spaces for the spins and the photons are disjoint. The square of the shifted auxiliary field contains a term independent of the field, $\mathcal{H}_{\gamma}^D=D(a^{\dagger}+a)^2$ with $D=\alpha_{\phi}^2/(2\beta) = \alpha^2/(2V_0)$, which has been shifted into the photon part of $\mathcal{H}_0$.

We are now in a position to trace over the spin degrees of freedom. This has been dealt with in detail elsewhere \cite{MckenzieStamp}; here we simply quote the result for the partition function in the random phase approximation
\begin{align}
\label{eq:Pfunction}
\frac{Z}{Z_{\mathcal{H}_0}Z_{\phi}} =  \biggr\langle\biggr\langle  T_{\tau}
\exp\biggr(\alpha_{\phi} \int_{\tau} \phi_0 (a^{\dagger}+a) \biggr)\biggr\rangle_{\phi}\biggr\rangle_{\gamma}
\end{align}
where $\langle \cdots \rangle_{\phi}$ is an average taken with respect to the free auxiliary field. 

Transforming to Matsubara frequency space
\begin{align}
\phi(i\omega_n) = \int_{\tau} e^{i\omega_n \tau} \phi(\tau),
\end{align}
the partition function of the free auxiliary field is
\begin{align}
Z_{\phi} = \int \mathcal{D}\phi \exp\biggr(-\frac{1}{2} \sum_{n,{\bf k}} 
(\mathcal{D}_{\phi}^0\bigr({\bf k},i\omega_n)\bigr)^{-1} |\phi_{\bf k}(i\omega_n)|^2\biggr),
\end{align}
where the free field propagator, $\mathcal{D}_{\phi}^0\bigr({\bf k},i\omega_n) = \langle |\phi_{\boldsymbol{k}}(i\omega_n)|^2\rangle_{\phi}$, is
\begin{align}
\mathcal{D}_{\phi}^0\bigr({\bf k},i\omega_n)
 = \frac{1}{1- V_{\boldsymbol{k}} \chi_0(i\omega_n)}
= 1+V_{\boldsymbol{k}} \chi(\boldsymbol{k},i\omega_n).
\end{align}
One may make use of the spectral decomposition of the dynamic susceptibility (equation (\ref{eq:specrep})) to determine the spectral representation of the free field propagator.

Beginning with a microscopic spin-photon Hamiltonian, we have developed an effective theory describing photons interacting with an auxiliary field which represents the collective magnetic excitations, or magnons, present in the system. We now turn to the photon component of the Hamiltonian, and make use of harmonic oscillator position and momentum operators to develop a path integral representation of the photonic degrees of freedom.

\subsection{Photon Hamiltonian}
\label{sec:Hgamma}

Considering a single photon mode, the photon Hamiltonian is given by
\begin{align}
\label{eq:Hgamma}
\mathcal{H}_{\gamma} = \omega_r \biggr(a^{\dagger} a + \frac{1}{2}\biggr)
- \lambda (a^{\dagger}+a) + D (a^{\dagger}+a)^2,
\end{align}
where
\begin{align}
\label{eq:D}
\lambda = \alpha \sqrt{N} \langle J^z \rangle_0
\qquad \text{and} \qquad
D = \frac{\alpha_{\phi}^2}{2\beta} = \frac{\alpha^2}{2V_0} .  
\end{align}
As discussed following equation (\ref{eq:MomentumCoupling}), the term linear in the photon operators is the instantaneous mean field Zeeman energy of the spins in the applied ac field, only now we are considering spins coupled to an effective photon position operator. The source of the diamagnetic term is the shift in the auxiliary field given in equation (\ref{eq:fieldshift}). Although the diamagnetic term is present in this intermediate stage of the development of the magnon-polariton theory, we find it does not play a role in the final expression for the magnon-polariton propagator given in Section \ref{sec:MPprop}.

We proceed by representing the photons with classical harmonic oscillator variables
\begin{align}
x = \sqrt{\frac{\hbar}{2m\omega}} (a^{\dagger}+a)
&& p = i \sqrt{\frac{\hbar m \omega}{2}} (a^{\dagger}-a).
\end{align}
In terms of these operators we have  ($\hbar,m=1$)
\begin{align}
\mathcal{H}_{\gamma} = \frac{p^2}{2} + \frac{1}{2} \omega_r^2 x^2 
- \sqrt{2\omega_r} \lambda x + 2 D \omega_r x^2.
\end{align}
The diamagnetic term shifts the oscillator frequency. In terms of the shifted variables
\begin{align}
\label{eq:freqshift}
\omega_{\gamma} = \omega_r \sqrt{1+\frac{4D}{\omega_r}} \qquad \text{and} \qquad
\lambda_{\gamma} = \lambda \biggr[1+\frac{4D}{\omega_r}\biggr]^{-\frac{1}{4}},
\end{align}
the photon Hamiltonian is 
\begin{align}
\mathcal{H}_{\gamma} = \frac{p^2}{2} + \frac{1}{2}  \omega_{\gamma}^2 x^2
-\sqrt{2 \omega_{\gamma}} \lambda_{\gamma} x.
\end{align}
The term linear in the position operator re-zeros the oscillator, and leads to a shift in its ground state energy
\begin{align}
\mathcal{H}_{\gamma} = \frac{p^2}{2} + \frac{1}{2}  \omega_{\gamma}^2 (x-x_0)^2
-\frac{1}{2}  \omega_{\gamma}^2 x_0^2,
\end{align}
where $x_0 = \sqrt{2\omega_{\gamma}} \lambda_{\gamma}/\omega_{\gamma}^2$. This linear shift of the oscillator will not affect the photon propagator. In terms of photonic quasiparticle operators which create and annihilate photons with energy $\omega_{\gamma}$, the photon Hamiltonian may be written
\begin{align}
\mathcal{H}_{\gamma} = \omega_{\gamma} \biggr(a_{\gamma}^{\dagger} a_{\gamma} + \frac{1}{2}\biggr)
-\frac{1}{2}  \omega_{\gamma}^2 x_0^2.
\end{align}
The shift in the ground state energy may be included with the ground state energy of the spins $E_{gs}$ (see the discussion following equation (\ref{eq:HMF})), and dropped from subsequent consideration.

In imaginary time, the propagator of the shifted photon modes is
\begin{align}
D_{\gamma}(\tau) = \bigr\langle T_{\tau} \bigr(a_{\gamma}^{\dagger}(\tau)+a_{\gamma}(\tau)\bigr) 
\bigr(a_{\gamma}^{\dagger}+a_{\gamma}\bigr) \bigr\rangle_{\gamma},
\end{align}
where the average $\langle \cdots \rangle_{\gamma}$ is taken with respect to $\mathcal{H}_{\gamma}$. One may express the partition function of the photon system in terms of a path integral over the harmonic oscillator position operator \cite{ShankarBook}
\begin{align}
Z_{\gamma} = \text{Tr}[e^{-\beta \mathcal{H}_{\gamma}}] = \int \mathcal{D}x \exp{\biggr[-\int_0^{\beta} 
\mathcal{L}_{\gamma}[\dot{x},x] d\tau\biggr]},
\end{align}
where $\mathcal{H}=\mathcal{L}$ in imaginary time, and the path integral is over the shifted harmonic oscillator variables. In the field theory, it is convenient to work with the dimensionless operator $x_{\gamma}=a_{\gamma}^{\dagger}+a_{\gamma}$. In Matsubara frequency space, the Euclidean action in terms of the dimensionless operator $x_{\gamma}$ is given by
\begin{align}
\label{eq:Lgamma}
\int_0^{\beta} \mathcal{L}_{\gamma}(\tau) d\tau = \frac{1}{2}\frac{\beta}{2\omega_{\gamma}}
\sum_n \bigr[-(i\omega_n)^2+\omega_{\gamma}^2\bigr] |x_{\gamma}(i\omega_n)|^2.
\end{align}
It follows that the photon propagator is given by
\begin{align} 
D_{\gamma}(i\omega_n)=\frac{2\omega_{\gamma}}{\beta} \frac{1}{\omega_{\gamma}^2-(i\omega_n)^2}.
\end{align}

We now have the free propagators of the magnon and photon systems, $\mathcal{D}_{\phi}^0\bigr({\bf k},i\omega_n)$ and $D_{\gamma}(i\omega_n)$. Equipped with these propagators, we may proceed to calculate the magnon-polariton propagator for the coupled magnon-photon system.

\subsection{Magnon-Polariton Propagator}
\label{sec:MPprop}

We have developed a path integral representation of the partition function for a quantum Ising system in a resonator. We return now to the partition function of the full system, given by equation (\ref{eq:Pfunction}). The non-interacting component of the partition function may be rewritten as $Z_{\mathcal{H}_0} = Z_{MF}Z_{\gamma}$, where $Z_{MF}$ yields the mean field free energy of the spins, and $Z_{\gamma}$ yields the free energy of the free photons. Although the mean field free energy of the spins has important thermodynamic consequences, it has no bearing on the magnon-polariton propagator and may be dropped from subsequent analysis. 

We define the magnon-polariton partition function by
\begin{align}
\label{eq:Zmp}
\frac{Z_{mp}}{Z_{\gamma}Z_{\phi}} =  \biggr\langle\biggr\langle  T_{\tau}
\exp\biggr(\alpha_{\phi} \int_{\tau} \phi_0 (a^{\dagger}+a) \biggr)
\biggr\rangle_{\phi}\biggr\rangle_{\gamma}.
\end{align}
We wish to determine the magnon-polariton propagator of the rescaled photon operators ($x_{\gamma} = a_{\gamma}^{\dagger}+a_{\gamma}$),
\begin{align}
D_{mp}^{\gamma}(\tau) = \bigr\langle T_{\tau} \bigr(a_{\gamma}^{\dagger}(\tau)+a_{\gamma}(\tau)\bigr) 
\bigr(a_{\gamma}^{\dagger}+a_{\gamma}\bigr) \bigr\rangle_{mp},
\end{align}
but in order to do so, we must re-express the interaction in terms of the photon operators $a_{\gamma}$. In terms of the dimensionless operator $x_{\gamma}=a_{\gamma}^{\dagger}+a_{\gamma}$, we find that 
\begin{align}
\alpha_{\phi} \int_{\tau} \phi_0 (a^{\dagger}+a) 
= \beta \alpha_{\gamma} \sum_n \phi(i\omega_n) x_{\gamma}(-i\omega_n)
\end{align}
where
\begin{align}
\label{eq:alpha}
\alpha_{\gamma} = \frac{\alpha_{\phi}}{\beta} \biggr[1+\frac{4D}{\omega_r}\biggr]^{-\frac{1}{4}}.
\end{align}
Note that if $D_{mp}$ is the propagator for the original photonic operators, $x=a^{\dagger}+a$, which create and annihilate photons with frequency $\omega_r$, we have $D_{mp}^{\gamma} = (\omega_{\gamma}/ \omega_r) D_{mp}$.

In order to calculate the magnon-polariton propagator, one may expand the interaction in (\ref{eq:Zmp}) and sum the resulting Dyson series. The exact result for the magnon-polariton propagator is
\begin{align}
D_{mp}^{\gamma}(i\omega_n) 
= \frac{1}{D_{\gamma}^{-1}(i\omega_n) - \beta^2 \alpha_{\gamma}^2 \mathcal{D}_{\phi}(i\omega_n)}.
\end{align}
Recall that the free field propagator may be written in terms of the dynamic susceptibility as
$\mathcal{D}_{\phi}^0\bigr(i\omega_n) = 1+V_0 \chi(i\omega_n)$, where $\chi(i\omega_n)$ is the zero wavevector component of the RPA susceptibility given by equation (\ref{eq:specrep}). This leads to
\begin{align}
D_{mp}^{\gamma}(i\omega_n) = -\frac{2\omega_{\gamma}}{\beta} \biggr[ 
\frac{1}{(i\omega_n)^2 - \omega_{c}^2 
+ (\alpha_c^2/\beta) \chi(i\omega_n)}\biggr],
\end{align}
where the effective frequency of the resonator and the effective coupling strength are now
\begin{align}
\omega_c^2 = \omega_{\gamma}^2 - 2\beta \alpha_{\gamma}^2 \omega_{\gamma}
\quad \text{and} \quad
\alpha_c^2 = 2 \beta^2 \alpha_{\gamma}^2 \omega_{\gamma} V_0.
\end{align}

The resonant frequency of the resonator is shifted by the diamagnetic response of the photons $\omega_r \rightarrow \omega_{\gamma}$ (equation (\ref{eq:freqshift})). The coupling between the photons and the auxiliary field again shifts the resonator frequency $\omega_{\gamma} \rightarrow \omega_c$. A short calculation shows that $\omega_c = \omega_r$, so the resonant photon frequency of the system is unchanged. This is as one might expect because the original spin-photon Hamiltonian does not contain a diamagnetic term. 

In terms of the original parameters of the spin-photon Hamiltonian, one may show that the rescaled coupling is $\alpha_c^2/\beta = \alpha^2 2\omega_r$. Using the fact that $D_{mp} = (\omega_r/\omega_{\gamma}) D_{mp}^{\gamma}$, we arrive at the magnon-polariton propagator of the original resonator photons ($x=a^{\dagger}+a$)
\begin{align}
\label{eq:MPprop2}
D_{mp}(z) = -\frac{2\omega_r}{\beta} \biggr[ 
\frac{1}{z^2 - \omega_r^2 
+ \alpha^2 2\omega_r \chi(z)}\biggr].
\end{align}
This propagator is a central result of the magnon-polariton theory. As discussed in Section \ref{sec:ResPhys}, it provides a primary connection between theoretical work and the experimentally measured resonator transmission function. Our result for the propagator includes the effects of counter-rotating terms which become important in the ultra-strong, or deep strong, coupling regimes \cite{Kockum}.

The dynamic susceptibility is given in equation (\ref{eq:specrep}). With $\chi_{el}=0$, one may write down an effective bosonic magnon-photon Hamiltonian describing the system 
\begin{align}
\label{eq:Hmp}
\mathcal{H}_{mp} = \omega_r &a^{\dagger} a  + \sum_{m} \omega_m b_m^{\dagger} b_m 
\\ \nonumber
&+ (a^{\dagger}+a) \sum_{m} g_m (b_m^{\dagger}+b_m).
\end{align}
In the absence of damping, the magnon-polariton propagator for the theory is given by (see appendix \ref{ap:CHO}) 
\begin{align}
\label{eq:MPprop3}
D_{mp}^{\mathcal{H}}(i\omega_n) 
= -\frac{2\omega_r}{\beta} \left[ \frac{1}{(i\omega_n)^2-\omega_r^2
-\sum_m \frac{4 g_m^2 \omega_m\omega_r}{(i\omega_n)^2-\omega_m^2}}\right].
\end{align}
Comparing with equation (\ref{eq:MPprop2}), we see that the coupling in the effective bosonic theory is
\begin{align}
\label{eq:coupling}
g_m^2 = \alpha^2 A_m.
\end{align} 
The propagator then satisfies $D_{mp}^{\mathcal{H}}(a^{\dagger},a) = D_{mp}(a^{\dagger},a)$. Recall that the spectral weights of the magnon modes scale like $A_m \sim 1/\omega_m$, so that the couplings will also scale like the inverse of the mode energies. The magnon mode energies, $\omega_m$, and the coupling strength, $g_m$, are temperature dependent due to the temperature dependence of the mean field, and the population factors which determine $A_m$.

One sees that the effective bosonic magnon-photon Hamiltonian captures the propagator of the original resonator photons coupled to the quantum Ising spins, apart from the contribution from the quasielastic diffusive pole. Therefore, when $\chi_{el}=0$, we are free to use the bosonic theory to describe the magnon-photon system. Note that there is no diamagnetic term in the effective bosonic Hamiltonian. In the Dicke model, one is dealing with mobile charged particles, and the diamagnetic term comes from squaring the canonical momentum of the charge carriers. As we are dealing with a fixed system of spins, no such term is expected.

In the development of the auxiliary field theory, we treated the magnetic fluctuations in the quantum Ising system in the RPA, and determined the exact magnon-polariton propagator within this approximation. In the rotating wave approximation (RWA), counter rotating terms in the effective bosonic Hamiltonian are dropped, leading to an approximate result for the magnon-polariton propagator \cite{Harder} (assuming $\chi_{el}=0$, and with $D_{mp}^{ret}(\omega) = \beta D_{mp}(z\rightarrow\omega+i0^+)$)
\begin{align}
D_{mp}^{ret}(\omega)\ = &\ \frac{1}{\omega - \omega_{mp}^- + i\Gamma_{mp}^-/2}
\\ \nonumber
&\qquad -\frac{1}{\omega + \omega_{mp}^+ + i\Gamma_{mp}^+/2} .
\end{align}
where
\begin{align}
\label{eq:RPA1}
\omega_{mp}^-(\omega) &= \omega_r  +
\sum_m \frac{g_m^2(\omega-\omega_m)}{(\omega-\omega_m)^2 + (\Gamma_m/2)^2}
\\ \nonumber
\omega_{mp}^+(\omega) &= \omega_r  -
\sum_m \frac{g_m^2(\omega+\omega_m)}{(\omega+\omega_m)^2 + (\Gamma_m/2)^2},
\end{align}
and
\begin{align}
\label{eq:RPA2}
\frac{\Gamma_{mp}^-(\omega)}{2} = \frac{\Gamma_r}{2} +
\sum_m \frac{g_m^2\Gamma_m/2}{(\omega-\omega_m)^2 + (\Gamma_m/2)^2}
\\ \nonumber
\frac{\Gamma_{mp}^+(\omega)}{2} = \frac{\Gamma_r}{2} + 
\sum_m \frac{g_m^2\Gamma_m/2}{(\omega+\omega_m)^2 + (\Gamma_m/2)^2}.
\end{align}
A phenomenological damping parameter $\Gamma_r$ has been included to account for any intrinsic damping of the resonator photons. As a coherent quantum Ising system is tuned through its critical point, the spectral weight of the soft mode diverges, as will the coupling of the soft mode to the resonator photons. When $g_m >> |\omega_r-\omega_m|$, one expects the RWA to break down, and it is necessary to make use of the full RPA magnon-polariton propagator to calculate resonator transmission.
 
\section{Discussion of Results}
\label{sec:DisRes}

We have developed an effective field theory, and an equivalent bosonic Hamiltonian, describing a quantum Ising system in a microwave resonator. The theory has been used to calculate the magnon-polariton propagator of the light-matter system.

In the Dicke and Hopfield models (see Appendix \ref{ap:Models}), the diamagnetic response of a light-matter system goes like the square of the coupling strength, $D \sim \alpha^2$, and the coupling strength varies like the square root of the atom or spin density, $\alpha \sim \rho^{\frac{1}{2}}$, as in equations (\ref{eq:BareCoupling}) and (\ref{eq:D}). With the diamagnetic term present, the effective resonator frequency diverges with the spin density (see Appendix \ref{ap:CHO}). This forestalls the superradiant quantum phase transition \cite{Rzazewski}, and leads to light-matter decoupling \cite{DeLiberato}. The situation here is different. Importantly, the effective Hamiltonian describing the magnon-photon system (equation (\ref{eq:Hmp})) does not contain a diamagnetic term. The magnon-photon coupling strength depends on the spectral weight of the relevant magnon mode (see equation (\ref{eq:coupling})), and may be tuned by the applied transverse field independently of the resonator frequency.

Consider a system with a single magnon mode. In the absence of damping, the upper and lower polariton modes follow from the poles of the magnon-polariton propagator (equation (\ref{eq:MPprop3}))
\begin{align}
\label{eq:wpm}
\omega_{\pm}^2 = \frac{\omega_r^2+\omega_m^2}{2} \pm 
\sqrt{\biggr(\frac{\omega_r^2-\omega_m^2}{2}\biggr)^2+4g_m^2\omega_r\omega_m}.
\end{align}
As the system is tuned through a quantum critical point, the spectral weight of the soft mode will diverge, as will the coupling $g_m^2 \sim A_m \sim 1/\omega_m \rightarrow \infty$. At the degeneracy point, $\omega_r=\omega_m$, there ought to be an avoided level crossing in the magnon-polariton spectrum
\begin{align}
\label{eq:wmpdeg}
\omega_{\pm} = \omega_r \sqrt{1 \pm 2  g_m/\omega_r} \qquad \text{if} \qquad
\omega_m=\omega_r,
\end{align}
or possibly a superradiant quantum phase transition if $g_m > \sqrt{\omega_m\omega_r}/2$. Recall from the discussion following equation (\ref{eq:coupling}) that $g_m$ and $A_m$ are temperature dependent, so the condition for superradiance is valid at finite temperatures. We have neglected dissipation and decoherence of the soft mode. The divergent spectral weight of the soft mode will lead to strong coupling to the resonator photons; it will also lead to strong coupling with bath degrees of freedom such as extraneous photons and phonons. 

Prior to a discussion of the damped magnon-polariton system, we provide a brief analysis of the propagator in the random phase approximation, and in mean field theory. In the random phase approximation, we capture the coupling between photons and collective spin excitations in the system. At the mean field level, we capture single ion excitations.

\subsection{Mean Field Theory}
\label{sec:MF}

In order to calculate the magnon-polariton propagator in the random phase approximation, an auxiliary field was introduced to account for the magnetic fluctuations. The resulting theory accounts for spins coupled to collective excitations in the material. In order to capture excitations at individual sites, a mean field theory (MF) is more appropriate. One may calculate the magnon-polariton propagator in MF theory without introducing the auxiliary field.

Our starting point for the MF calculation is equation (\ref{eq:PF}), where $\mathcal{H}'$ is now
\begin{align}
\mathcal{H}' = - \alpha (a^{\dagger}+a) \frac{1}{\sqrt{N}} \sum_i \delta J_i^z.
\end{align}
We have dropped the interactions between the fluctuations of the spins about their MF. The photon Hamiltonian is given by
\begin{align}
\mathcal{H}_{\gamma} = \omega_r \biggr(a^{\dagger} a + \frac{1}{2}\biggr)
- \lambda (a^{\dagger}+a).
\end{align}
At the MF level, there is no diamagnetic term in the photon Hamiltonian. The diamagnetic term came from a shift in the auxiliary field in the RPA theory.

One may introduce harmonic oscillator variables, as in Section \ref{sec:Hgamma}, to obtain an effective action for the photons. The result is the same as in equation (\ref{eq:Lgamma}), with $\omega_{\gamma}$ replaced with the resonator frequency $\omega_r$. Recall that the shift in the photon frequencies ($\omega_r \rightarrow \omega_{\gamma}$) came from the diamagnetic term in the photon Hamiltonian, which is not present in the MF theory. 

The resulting MF magnon-polariton partition function is (recall $\int_{\tau} \equiv \int_0^{\beta} d\tau / \beta$)
\begin{align}
\frac{Z_{mp}^{MF}}{Z_{\gamma}} =  \biggr\langle\biggr\langle  T_{\tau}
\exp\biggr(\beta \alpha \int_{\tau} x \delta J_0^z \biggr)\biggr\rangle_{s}\biggr\rangle_{\gamma},
\end{align}
where $x=a^{\dagger}+a$, and $\delta J_0^z$ is the zero wavevector component of the electronic spin operators. One may perform a cumulant expansion and trace over the microscopic spin degrees of freedom \cite{MckenzieStamp}. The average over the spins $\langle \cdots \rangle_s$ is taken with respect to the MF spin Hamiltonian. We have dropped $Z_{MF}$ from $Z_{mp}^{MF}$ because the mean field partition function of the spins plays no further role in determining the magnon-polariton propagator. 

Truncating the result of the cumulant expansion at the RPA (or Gaussian) level, one finds
\begin{align}
\frac{Z_{mp}^{MF}}{Z_{\gamma}} =  \biggr\langle  \exp\biggr(
\frac{\beta \alpha^{2}}{2} \sum_n \chi_0(i\omega_n) |x(i\omega_n)|^2\biggr) \biggr\rangle_{\gamma},
\end{align}
and the resulting mean field magnon-polariton propagator is
\begin{align}
D_{mp}^{MF}(z) = -\frac{2\omega_r}{\beta} \biggr[ 
\frac{1}{z^2-\omega_r^2+\alpha^2 2\omega_r \chi_0(z)}\biggr].
\end{align}
This result may have easily been anticipated from equation (\ref{eq:MPprop2}). Writing the RPA susceptibility as a Born series we have
\begin{align}
\chi = \chi_0 + \chi_0 V_0 \chi_0 + \chi_0 V_0 \chi_0 V_0 \chi_0 + \cdots 
\end{align}
Truncating the series after the first term leads to the MF result involving light scattering from individual ions. Summing the full series leads to the RPA result which describes light coupled to collective modes of the system. We have derived the MF result here to demonstrate the use of the magnon-polariton theory at the MF level, where the introduction of the auxiliary field is unnecessary.

Using the spectral decomposition of the MF propagator, which follows from equation (\ref{eq:chi0}), and neglecting the quasielastic diffusive pole, one may write the propagator as
\begin{align}
\label{eq:DmpMF}
D_{mp}^{MF}(z) = -\frac{2\omega_r}{\beta}\left[\frac{1}{z^2-\omega_r^2 -\sum_{n>m} 
\frac{4 g_{mn}^2 E_{nm} \omega_r}{z^2 - E_{nm}^2}} \right].
\end{align}
The coupling strength is $g_{mn}^2 = \alpha^2 a_{mn}$, where $a_{mn} = |c_{mn}|^2 p_{mn}$ is the spectral weight of the MF transition between states $n$ and $m$.

In a resonator experiment, one expects both single ion excitations, and collective modes. The eigenstates of the collective modes may involve quantum coherent superpositions of many different single ion eigenstates. As the system is subject to decoherence, the collective mode behavior may give way to single ion excitations. As will be demonstrated for the LiHoF$_4$ system, the relative strengths of the single ion excitations and the collective modes can be compared by tuning their respective spectral weights.

\subsection{Random Phase Approximation}
\label{sec:RPA}

In the absence of damping, we obtain the magnon-polariton propagator in the random phase approximation. A spectral decomposition of the magnon-polariton propagator may be obtained making use of equation (\ref{eq:MPprop3}), which we write as
\begin{align}
\beta D_{mp}(z)\biggr|_{RPA} = -\frac{P(z)}{Q(z)},
\end{align}
where
\begin{align}
P(z) = 2\omega_r \prod_m (z^2-\omega_m^2),
\end{align}
and
\begin{align}
\label{eq:resSpec}
Q(z) &= (z^2-\omega_r^2)\prod_m (z^2-\omega_m^2) 
\\ \nonumber
& \qquad - \sum_m 4g_m^2 \omega_m\omega_r \prod_{m'\neq m} (z^2-\omega_{m'}^2).
\end{align}
The magnon-polariton modes $ \{ \omega_p \} $ follow from the zeros of $Q(z)$, which we may rewrite as $Q(z) = \prod_p (z^2-\omega_p^2)$. The spectral decomposition of the propagator is then
\begin{align}
\beta D_{mp}(z)\biggr|_{RPA} = \sum_p \frac{A_p 2\omega_p}{\omega_p^2-z^2},
\end{align}
with
\begin{align}
\label{eq:ResSpecWeight}
A_p = \frac{\omega_r \prod_m (\omega_p^2-\omega_m^2)}{\omega_p \prod_{q \neq p}(\omega_p^2-\omega_q^2)}.
\end{align}
We see that the magnon-polariton spectral weights scale like the inverse of the mode energy, $A_p \sim 1/\omega_p$. This RPA expression will be used to calculate the modes of LiHoF$_4$ in a microwave resonator.

\subsection{Damped Magnon-Polariton Propagator}
\label{sec:DampedProp}

We have developed a theory of magnon-polaritons in quantum Ising systems, and discussed the resulting propagator in the random phase approximation and in mean field theory. We find that the magnon-photon coupling strength depends on the spectral weight of the relevant magnon mode. As a system is tuned through its quantum critical point, the divergent spectral weight of the soft mode leads to deep strong coupling between the soft mode and the resonator photons. As no diamagnetic term is present in the theory, one expects this to lead to a superradiant quantum phase transition. However, this neglects the effects of damping and decoherence due to the system's coupling to its environment, which we discuss here.

Coupled light-matter systems, and associated quantum technologies, are generating considerable excitement \cite{Kockum, LQReview,HuRoadmap,BhoiKim}. Of course, in any real-world scenario, one must consider the impact of the environment on the sytem of interest. For recent research on this topic see \cite{CorteseDeLiberato} and references therein. In this work, we do not explore the full complexity of the memory effects, dissipation, and decoherence expected when a polaritonic system is coupled to a bath, or baths; rather, we introduce phenomenological parameters that may account for damping and decoherence in the magnon-polariton theory at a basic level. The results are then compared to experimental data in Section \ref{sec:Expt}.

We assume ohmic (frequency independent) damping of the magnon modes, in which case the damped retarded magnon-polariton propagator may be written ($D_{mp}^{ret}(\omega) = \beta D_{mp}(z \rightarrow \omega+i0^+)$)
\begin{align}
\label{eq:dampedprop}
D_{mp}^{ret}(\omega) 
= \frac{-2\omega_r}{\omega^2-\omega_{mp}^2 + i\omega \Gamma_{mp}} ,
\end{align}
where from equation (\ref{eq:MPprop2})
\begin{align}
\omega_{mp}^2 = \omega_r^2 + (\Gamma_r/2)^2 - 2 \alpha^2 \omega_r \chi'(\omega),
\end{align}
and
\begin{align}
\omega \Gamma_{mp} = \omega \Gamma_r + 2 \alpha^2 \omega_r \chi''(\omega).
\end{align}
A factor of $\Gamma_r$ has been included to account for any intrinsic damping of the resonator mode. The $\Gamma_r$ term in the expression for $\omega_{mp}$ is a counterterm which eliminates a shift in the resonator frequency due to its damping (see the discussion in Section \ref{sec:DS}). The reactive and absorptive components of the dynamic susceptibility are given in equations (\ref{eq:chiprime}) and (\ref{eq:chiprime2}). The magnon damping functions $\{ \Gamma_m \}$, are assumed to be frequency independent, although they will vary with the transverse field. The magnon-polariton propagator can be viewed as a damped photon propagator, but the magnon ``bath" leads to a frequency dependent damping function, and a complex set of magnon-polariton modes that follow from the zeros of $\omega^2-\omega_{mp}^2(\omega)$, or equivalently $\omega_{mp}(\omega_p) = \omega_p$. 

Consider a system with a single magnon mode for which the polariton modes follow from the real part of
\begin{align}
\omega_p^2 = \omega_r^2 - (\omega_p + i\Gamma_r/2)^2 
- \frac{4g_m^2\omega_r\omega_m}{\omega_m^2-(\omega_p+i\Gamma_m/2)^2}.
\end{align}
When the damping is weak, we recover the upper and lower polariton modes given in equation (\ref{eq:wpm}). A superradiant phase transition will occur if the coupling strength is sufficiently strong to drive the lower polariton mode to zero. In the damped system, the condition for superradiance is
\begin{align}
\label{eq:gmDamped}
g_m > \frac{\sqrt{\omega_m\omega_r}}{2} \biggr[\biggr(1+\frac{\Gamma_r^2}{4\omega_r^2}\biggr)
\biggr(1+\frac{\Gamma_m^2}{4\omega_m^2}\biggr)\biggr]^{\frac{1}{2}}.
\end{align}
In the absence of damping and decoherence, if a magnon mode softens to zero, the magnon-polariton system will always be driven into a superradiant phase (recall $g_m \rightarrow \infty$ as $\omega_m \rightarrow 0$). With damping present, the divergence of $g_m$ may be matched by a divergence on the right hand side of equation (\ref{eq:gmDamped}) preventing the lower polariton mode from dropping to zero. Furthermore, if the constituent spins making up the soft mode are subject to decoherence, one expects a reduction in its spectral weight, and hence a reduction in the coupling strength $g_m$. This may lead to weak coupling and prevent superradiance. We will elaborate on this point in Section \ref{sec:Expt}.  

In a damped system comprised by multiple magnon modes, the magnon-polariton mode and linewidth equations are
\begin{widetext}
\begin{align}
\label{eq:wmp}
\omega_{mp}^2 = \omega_r^2 + \biggr(\frac{\Gamma_r}{2}\biggr)^2
+ \sum_m\frac{2 g_m^2 \omega_r (\omega-\omega_m)}{(\omega-\omega_m)^2 + (\Gamma_m/2)^2}
- \sum_m\frac{2 g_m^2 \omega_r (\omega+\omega_m)}{(\omega+\omega_m)^2 + (\Gamma_m/2)^2}
- 2 g_0^2 \omega_r \frac{(\Gamma_0/2)^2}{\omega^2-(\Gamma_0/2)^2}
\end{align}
and
\begin{align}
\label{eq:Gammamp}
\omega \Gamma_{mp} = \omega \Gamma_r
+ \sum_m \frac{g_m^2 \omega_r \Gamma_m}{(\omega-\omega_m)^2 + (\Gamma_m/2)^2}
- \sum_m \frac{g_m^2 \omega_r \Gamma_m}{(\omega+\omega_m)^2 + (\Gamma_m/2)^2}
+2 g_0^2 \omega_r \frac{\omega \Gamma_0/2}{\omega^2 +(\Gamma_0/2)^2}.
\end{align}
\end{widetext}
Note that the coupling to the zero mode, defined by $g_0^2 \equiv \alpha^2 \chi_{el}$, will decay exponentially with temperature, and vanish in the paramagnetic phase of the system. We will drop this mode from subsequent consideration. One may compare the results for $\omega_{mp}$ and $\Gamma_{mp}$ with the RWA results given in equations (\ref{eq:RPA1}) and (\ref{eq:RPA2}). As previously noted, as a coherent quantum Ising system is tuned through its critical point, one expects the RWA results to break down.

Assuming that $g_m$ is sufficiently weak, or $\Gamma_m$ is sufficiently strong, to prevent superradiance, one finds the damping of the polariton mode at resonance ($\omega_p = \omega_m$) to be
\begin{align}
\Gamma_p = \Gamma_{mp}(\omega_p) \approx \Gamma_r \biggr[1+C\frac{\omega_r}{\omega_p}\biggr],
\end{align}
where the cooperativity of the system is $C=4g_m^2/(\Gamma_m \Gamma_r)$. Recall that $\omega_p$ is the polariton mode energy, $\omega_m$ is the magnon mode energy, and $\omega_r$ is the bare resonance frequency of the resonator. From inspection of equation (\ref{eq:Gammamp}), we see that as the soft magnon mode is tuned through $\omega_p$, a resonance is expected to appear in the linewidth of the polariton mode. The form and magnitude of the resonance will depend on how $g_m$ and $\Gamma_m$ vary with the transverse field.

The collective magnon modes, in particular the soft mode, are entangled many-body eigenstates of the spin system. The quantum coherence of the collective magnon modes is not easily accounted for by the theory. When the quantum coherent superposition of spins comprising a particular magnon mode are in contact with their environment, one expects the superposition of spin states to give way to a classical mixture of spin states. In our analysis of the LiHoF$_4$ system below, we account for this decoherence by transferring spectral weight from the collective RPA excitations to the single ion excitation spectrum. This leads to mixed single ion and collective mode transmission in the magnon-polariton propagator.

With dissipation and decoherence present, in the limit $g_m/\Gamma_m \rightarrow 0$, we have $\omega_{mp} = \omega_r$ and $\Gamma_{mp} = \Gamma_r$. The resonator shows no evidence of the magnon modes. We note, however, that this is distinct from the light-matter decoupling discussed by De Liberato \cite{DeLiberato}. In light-matter decoupling, the diamagnetic response of the system localizes the photon modes away from the matter modes and shifts the frequency of the photons, so that the polaritonic quasiparticle operators have a distinct light or matter character. The diamagnetic term is absent in the magnon-polariton theory, and the environment is an additional feature that may prevent superradiance.

\section{Comparison to Experiment}
\label{sec:Expt}


So far, our analysis has been theoretical. In order to have confidence in the results, one must compare theoretical work to experimental data. We do so here by comparing the magnon-polariton theory to transmission spectra of LiHoF$_4$ in loop gap microwave resonators \cite{Libersky, LiberskySM}.

Consider the low temperature effective Hamiltonian of the LiHoF$_4$ system \cite{Chakraborty, Tabei, MckenzieStamp}
\begin{align}
\label{eq:LiHo}
\mathcal{H}_{eff} =  
- \frac{C_{zz}^2}{2} \sum_{i \neq j} V_{ij} \tau_{i}^{z}\tau_{j}^{z}
- \frac{\Delta}{2} \sum_{i} \tau_{i}^{x} + \mathcal{H}_{hyp},
\end{align}
where the interaction contains a dipolar component, and a weaker antiferromagnetic component
\begin{align}
\label{eq:V}
V_{ij} = J_D D_{ij}^{zz} - J_{nn}.
\end{align}
In what follows, we assume a LiHoF$_4$ sample with zero demagnetization field, consistent with a needle shaped sample, or a striped domain pattern in which the demagnetization field in the bulk of the sample averages to zero (see Section IIB of the supplement to reference \cite{LiberskySM} for more details). The eigenstates of the $J=8$ holmium spins are mixed and split by the crystal electric field and an applied transverse field. The $\{\tau_i^{\mu}\}$ are Pauli operators describing the two lowest electronic spin states , and $C_{zz}(B_x)$ is a truncation parameter which depends on the applied transverse field, as does the effective transverse field, $\Delta(B_x)$, which splits the energies of the two lowest electronic spin eigenstates. The truncated longitudinal holmium electronic spin operator is $J^z = C_{zz} \tau^z$. The hyperfine component of the Hamiltonian contains the coupling of each effective spin-$1/2$ operator to its $I=7/2$ nucleus. This splits the single ion Hamiltonian into $16$ electronuclear levels, all of which can be accomodated using our formalism. Further details of the LiHoF$_4$ system are discussed in Appendix \ref{ap:LiHo}.

The spin-photon interaction is assumed to be
\begin{align}
\mathcal{H}_{int} =  -\alpha (a^{\dagger}+a) \delta J_0^{z},
\end{align}
where $\delta J_0^z = C_{zz} \delta \tau_0^z$ is the $k=0$ wavevector component of the longitudinal spin fluctuation operator ($\delta J_0^z = J_0^z - \langle J^z \rangle_{MF}$) in Fourier space, and the coupling constant is (see equation \ref{eq:BareCoupling})
\begin{align}
\label{eq:BareCoupling2}
\alpha = \eta \sqrt{2\pi} \sqrt{\hbar \omega_r} \sqrt{\rho J_D}.
\end{align}
In LiHoF$_4$ we have four spins in each unit cell having volume $V_{cell} = 2.88 \times 10^{-28}m^3$, to give a total number of spins $N=4V_{sample}/V_{cell}$. The spin density is $\rho=4/V_{cell}=1.39 \times 10^{28} m^{-3}$, which is about 3.3 times the value in YIG, and the dipolar energy per unit cell is $\rho J_D = 13.52mK = 282MHz$. The filling factor $\eta$ is left as a free parameter which depends on details of the resonator.

Consider, as an example, an $\omega_r/(2\pi)=1$GHz applied ac field. In temperature units, we have $\hbar \omega_r/k_B = 48mK$. Plugging in the numbers, we find the coupling at 1GHz to be
\begin{align}
\alpha \big|_{1GHz} \approx \eta \times 64mK = \eta \times 1.33GHz.
\end{align}
Using this value as a reference, the coupling for any given frequency (in GHz) is given by
\begin{align}
\alpha (f) \approx \eta \sqrt{f/f_0} \times 1.33 GHz,
\end{align}
where $f_0=1GHz$ is the reference frequency.

The resonator transmission function is given by equation (\ref{eq:ResTrans}). It follows from equation (\ref{eq:dampedprop}) that
\begin{align}
|S_{21}|^2 \propto \text{Im}[D_{mp}^{ret}] =
\frac{2\omega\omega_r\Gamma_{mp}}{(\omega^2-\omega_{mp}^2)^2 + (\omega\Gamma_{mp})^2}.
\end{align}
Without knowledge of the proportionality constant, one cannot obtain $\Gamma_{mp}$ from the amplitude and phase of the transmission function; however, one may still obtain the magnon-polariton modes, and compare qualitative features of their linewidths with theoretical results.

\begin{figure}[htp]
\centering
\includegraphics[width=8cm]{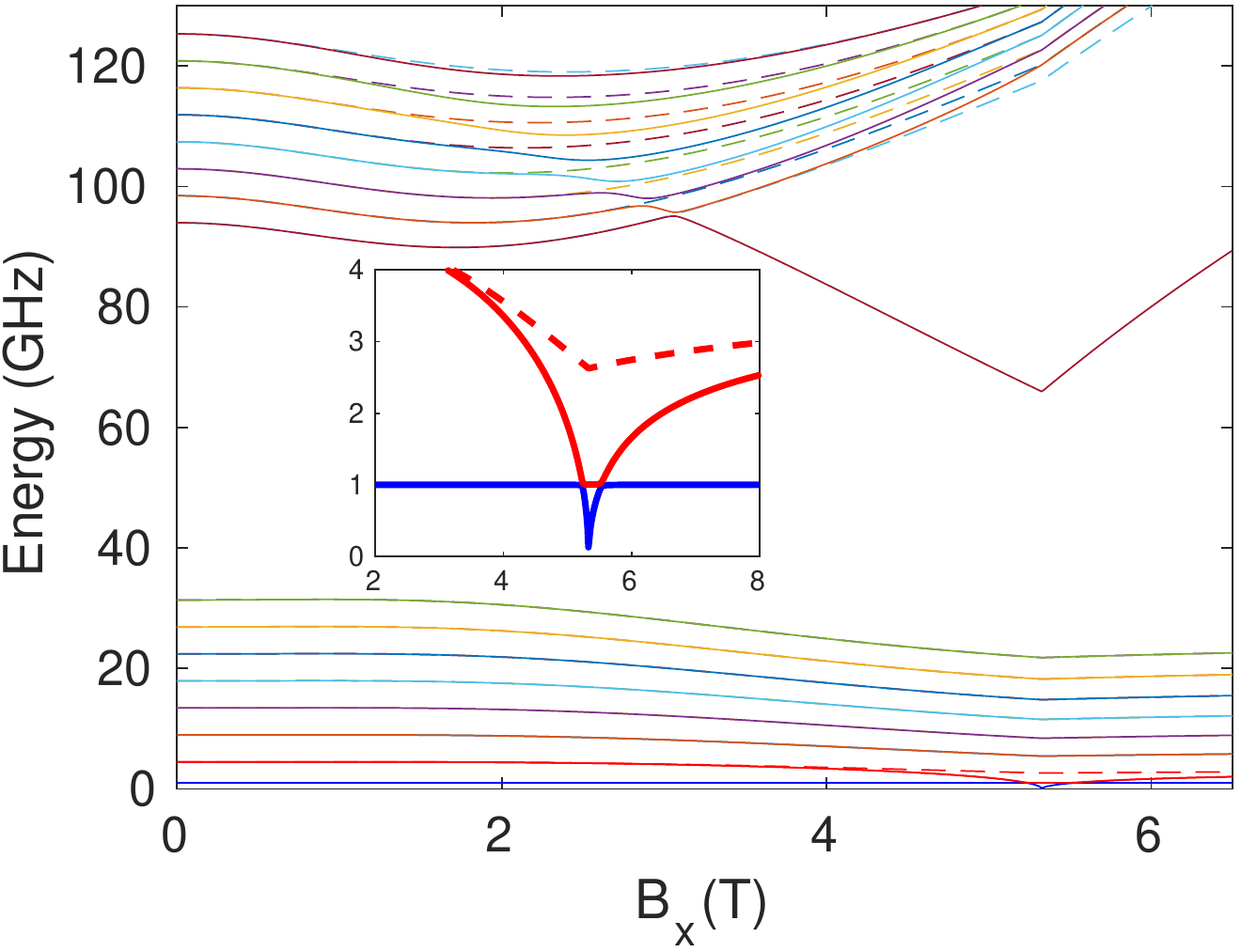}
\caption{Modes of LiHoF$_4$, at zero temperature, in a 1GHz resonator with a filling factor of $\eta=0.01$. We assume the average demagnetization field in the sample is zero. Due to the weak coupling, the RPA modes of the resonator are much the same as the RPA modes of LiHoF$_4$, with an additional mode at the resonator frequency. The MF modes of the LiHoF$_4$ system are shown as dashed lines for comparison. The mode showing significant softening in the upper band of energy levels has dominant spectral weight. This mode has been measured in neutron scattering experiments \cite{Ronnow2}. In the inset, we see the lowest energy electronuclear mode soften to zero at the quantum critical point. A similar figure showing the electronuclear modes of LiHoF$_4$, and their spectral weights, is provided in reference \cite{MckenzieStamp}.}
\label{fig:ModesWeakCoupling}
\end{figure}

If the coupling between the resonator photons and the magnetic excitations is weak, the modes of the resonator will differ little from the modes of LiHoF$_4$, apart from the appearance of an additional mode corresponding to the resonator frequency. In Figure \ref{fig:ModesWeakCoupling}, we illustrate the RPA transmission spectrum of a needle shaped sample of LiHoF$_4$ in a $1GHz$ resonator, at zero temperature, with a filling factor of $\eta=0.01$, along with the MF modes of LiHoF$_4$. The low energy RPA modes of the resonator differ little from the RPA modes of LiHoF$_4$, apart from the addition of the resonator mode. In the upper band of excitations, we see the gapped electronic mode which has been measured in neutron scattering experiments \cite{Ronnow2}. A comparison of spectral weights determined by equation (\ref{eq:ResSpecWeight}) shows that, under weak coupling, the RPA transmission spectrum of the resonator is dominated by the resonator mode. 

\begin{figure}
\centering
\begin{minipage}{.23\textwidth}
    \centering
    \includegraphics[width=1\linewidth,trim={0.1cm 0.2cm 0.5cm 0.5cm},clip]{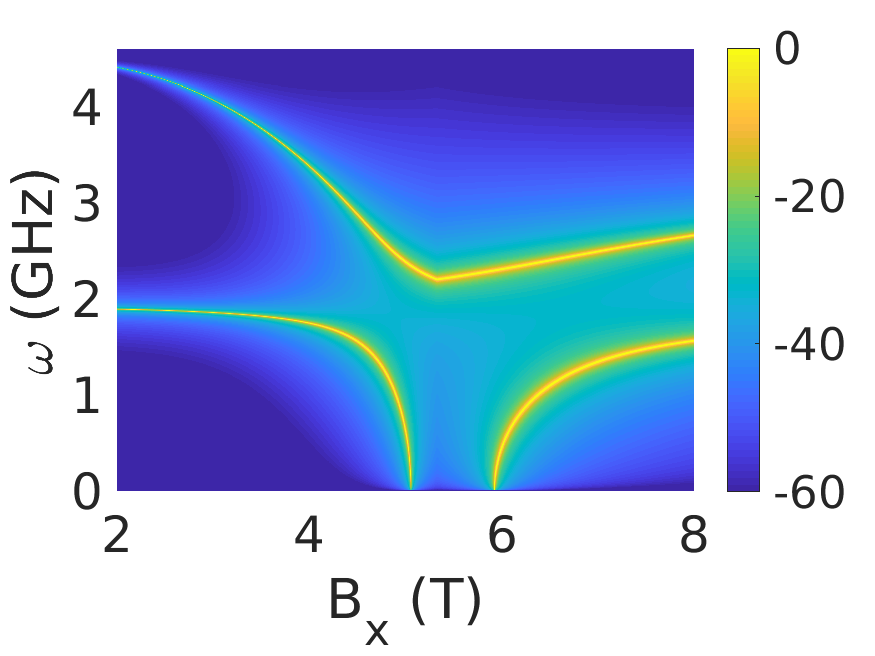}  
\end{minipage}
\begin{minipage}{.23\textwidth}
    \centering
    \includegraphics[width=1\linewidth,trim={0.05cm 0.2cm 0.3cm 0.5cm},clip]{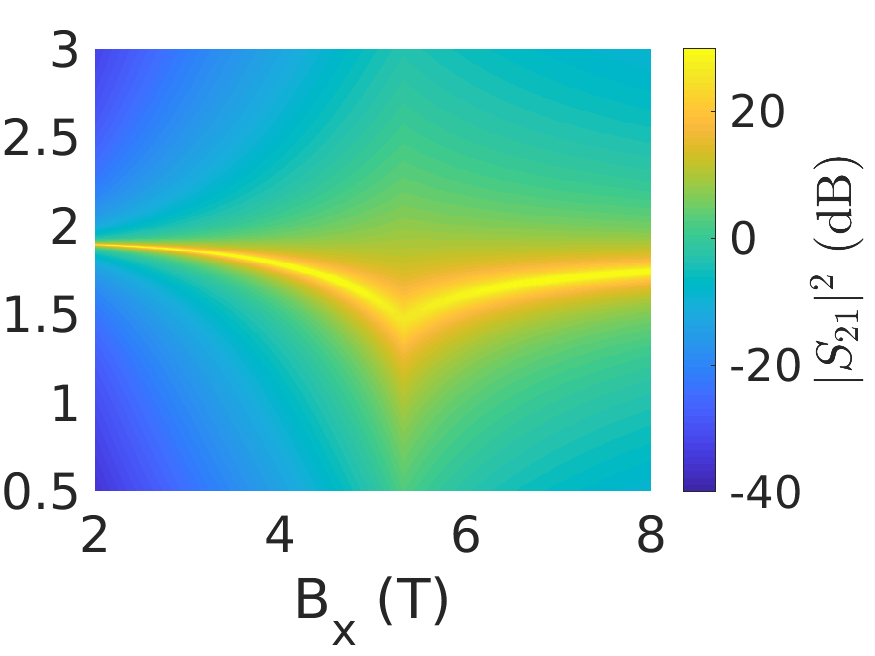}  
\end{minipage}
\caption{Damped RPA transmission function of LiHoF$_4$ in a 1.9GHZ microwave resonator at zero temperature. We consider a sample of LiHoF$_4$ in which the average demagnetization field is zero, and assume a filling factor of $\eta=0.25$. In the upper left hand figure, the dampings of the magnon modes and the resonator mode are $\Gamma_m = 1\mu K =20.837kHz$ and $\Gamma_r=1nK=20.837Hz$, respectively. With this weak damping, the system is driven into a superradiant phase. On the right, the damping of the soft mode has been increased to $\Gamma_{m=1} = 0.5K=10.419GHz$ which stops the lower polariton mode from softening to zero, preventing the superradiant phase transition. The upper polariton mode is attenuated to the point where it is no longer visible in the transmission spectrum.}
\label{fig:DampedTransmission}
\end{figure}

\begin{figure}
\centering
\begin{minipage}{.23\textwidth}
    \centering
    \includegraphics[width=1\linewidth,trim={0.2cm 0.1cm 0.05cm 0.1cm},clip]{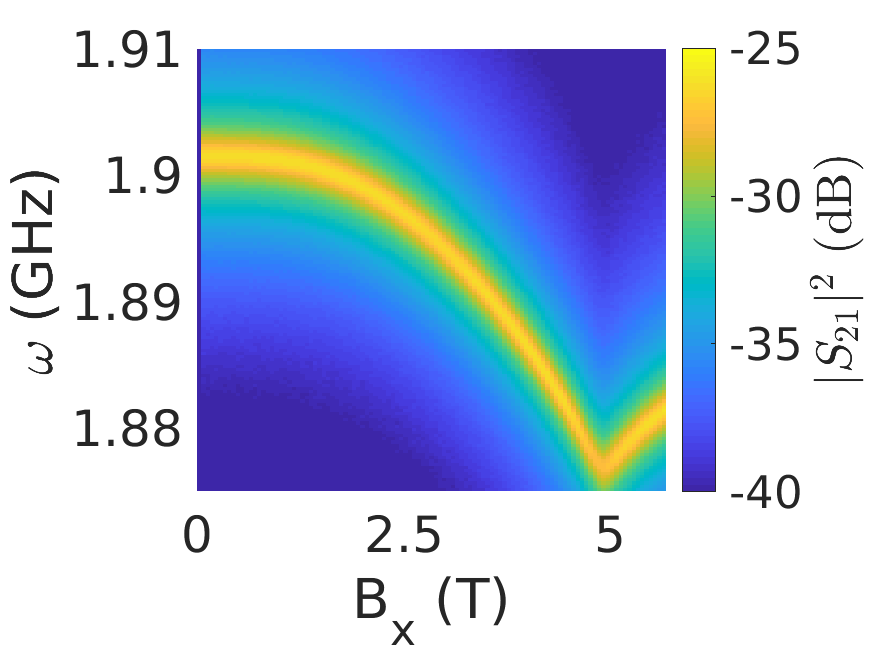}  
\end{minipage}
\begin{minipage}{.23\textwidth}
    \centering
    \includegraphics[width=1\linewidth,trim={0.1cm 0.1cm 0.05cm 0.1cm},clip]{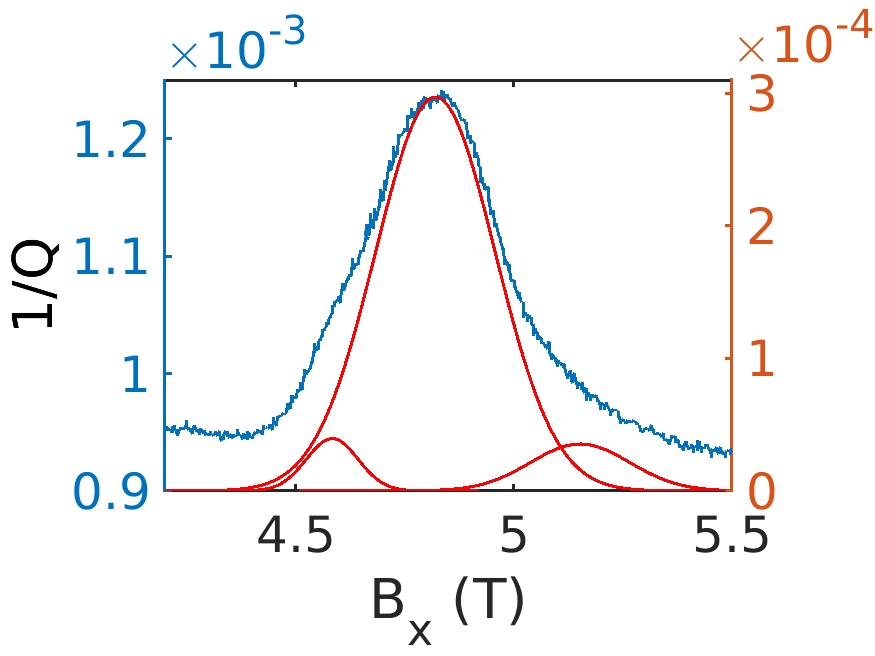}  
\end{minipage}
\caption{Measured transmission function of LiHoF$_4$ in a 1.9GHZ microwave resonator. The inverse quality factor of the resonator mode is shown on the right. The measured value of $1/Q$ (blue) has been decomposed into the sum of three Gaussian peaks (red). The central peak corresponds to absorption at the phase transition. The satellite peaks to either side of the central peak occur where the soft mode is degenerate with the resonator mode.}
\label{fig:ExptTransmission}
\end{figure}

\begin{figure}
\centering
\begin{minipage}{.23\textwidth}
    \centering
    \includegraphics[width=1\linewidth,trim={0.2cm 0.2cm 0.5cm 0.5cm},clip]{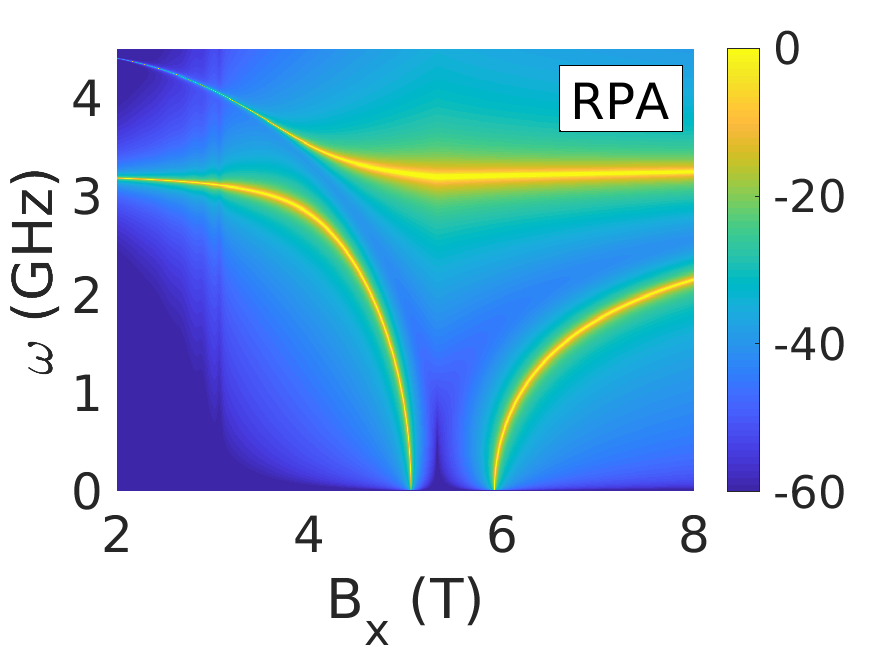}  
\end{minipage}
\begin{minipage}{.23\textwidth}
    \centering
    \includegraphics[width=1\linewidth,trim={0.1cm 0.2cm 0.3cm 0.5cm},clip]{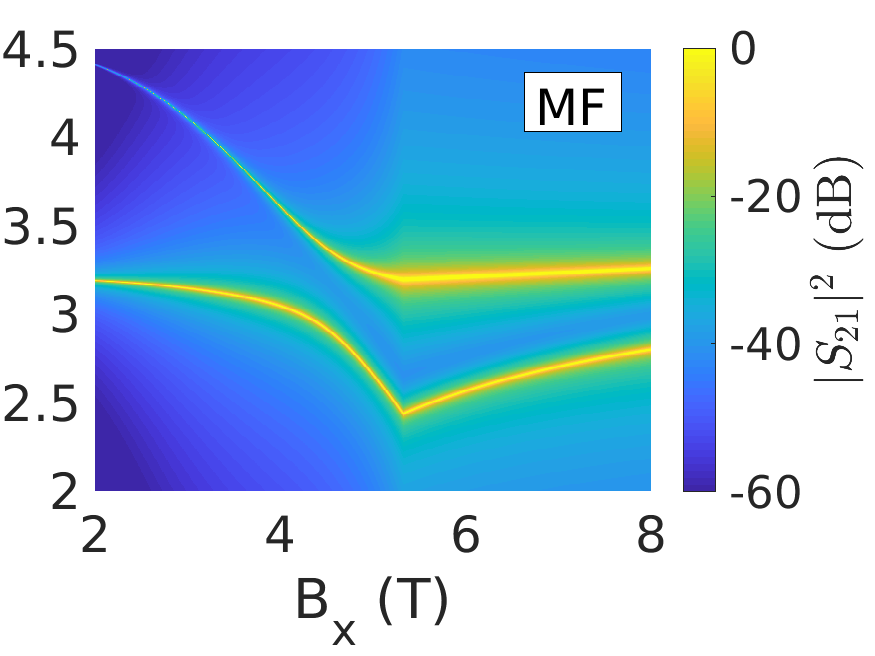}  
\end{minipage}
\caption{Damped transmission function of LiHoF$_4$ in a $3.2GHz$ microwave resonator with a filling factor of $\eta=0.25$ in the RPA (left), and in MF theory (right). We consider the zero temperature transmission of a LiHoF$_4$ sample with zero average demagnetization field. The damping parameters are given by $\Gamma_m \propto A_m \omega_m^2$ (similarly for the MF modes). The proportionality constant is chosen so that the damping parameters are roughly in line with what one expects for spin vacancies in diamond (see text). In the RPA (left), the damping is insufficient to prevent superradiance in the system due to the divergent spectral weight of the magnon soft mode. The spectral weight carried by the lowest MF mode does not diverge, and the resultant coupling strength is not strong enough to cause superradiance in the single ion resonator spectrum (right). In a system subject to decoherence, one expects to see both single ion and collective mode transmission.}
\label{fig:DampedModes}
\end{figure}

\begin{figure}[htp]
\centering
\begin{minipage}{.23\textwidth}
    \centering
     \includegraphics[width=1\linewidth,trim={0.1cm 0.1cm 0.5cm 0.8cm},clip]{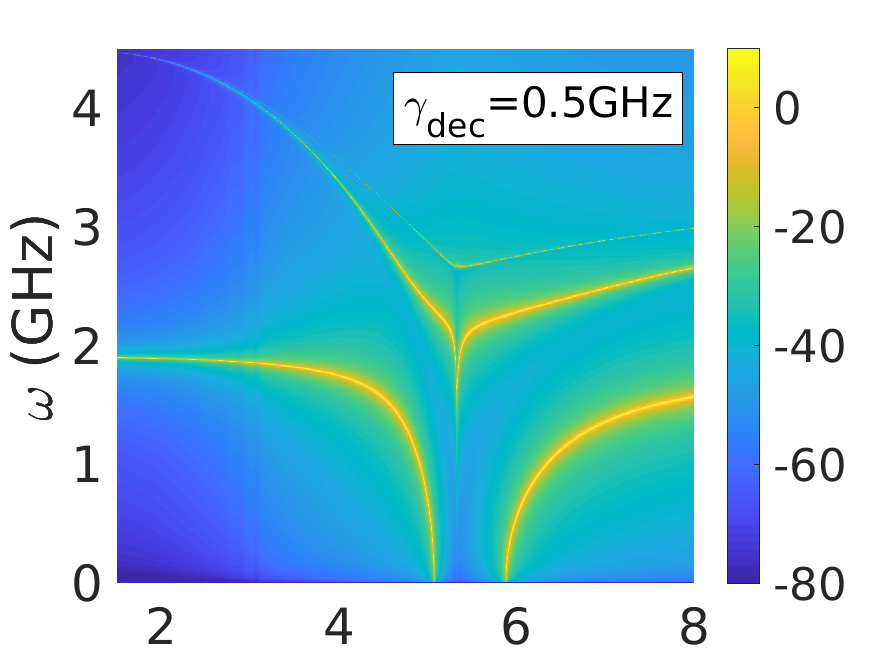} 
\end{minipage}
\begin{minipage}{.23\textwidth}
    \centering
     \includegraphics[width=1\linewidth,trim={0.8cm 0.1cm 0.3cm 0.6cm},clip]{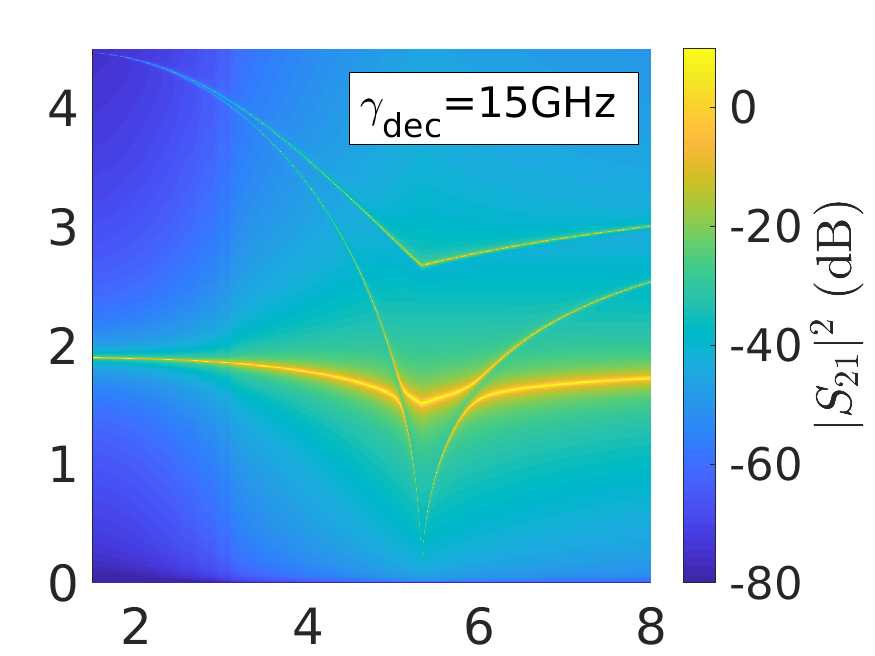} 
\end{minipage}
\begin{minipage}{.23\textwidth}
    \centering
     \includegraphics[width=1\linewidth,trim={0.1cm 0.2cm 0.5cm 0.5cm},clip]{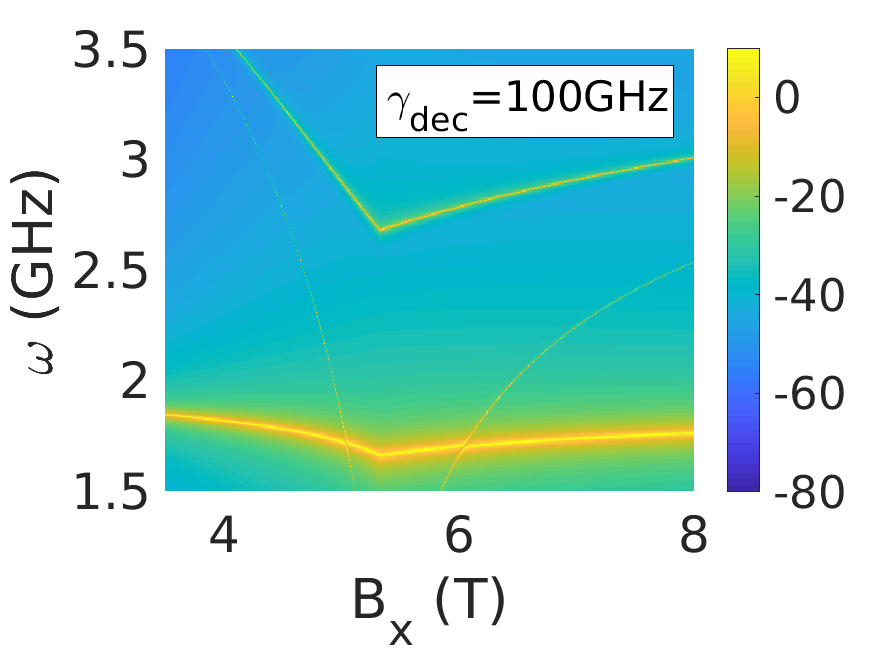} 
\end{minipage}
\begin{minipage}{.23\textwidth}
    \centering
     \includegraphics[width=1\linewidth,trim={0.15cm 0.2cm 0.3cm 0.5cm},clip]{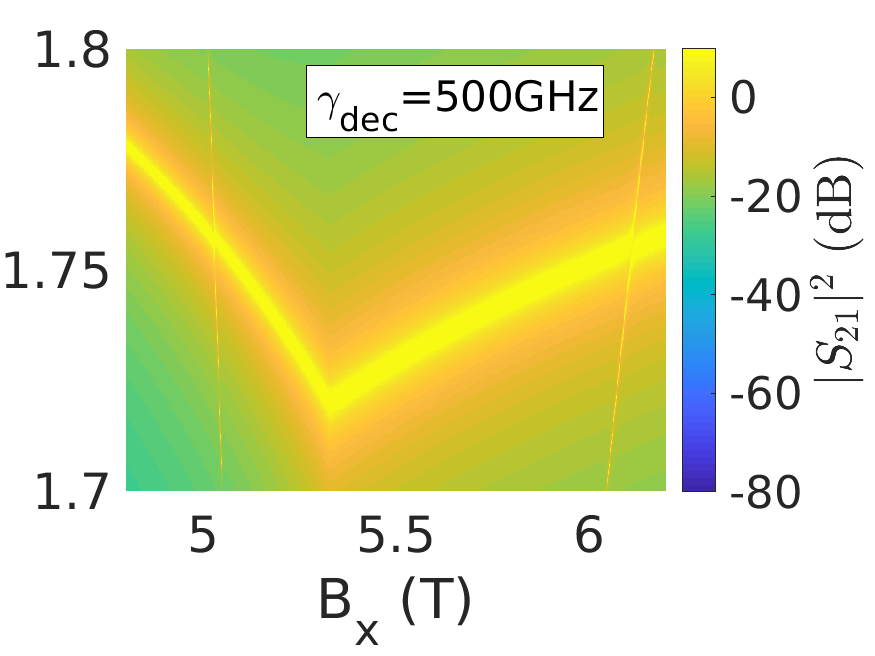} 
\end{minipage}
\caption{Mixed single ion and collective mode transmission of a LiHoF$_4$ sample in a 1.9GHz resonator at zero temperature. The filling factor is set to $\eta=0.25$ and we assume the average demagnetization field is zero. The damping parameters are chosen to be in line with what one might expect for spin vacancies in diamond. In the upper left figure the decoherence factor is set to $\gamma_{dec}=0.5GHz$. The sharp dip in the upper polariton mode occurs where the soft mode crosses $\gamma_{dec}$. In the upper right figure the decoherence factor has been increased to $\gamma_{dec}=15GHz$, which is sufficient to prevent superradiance. Further increasing the decoherence factor attenuates the soft mode and closes the avoided level crossing in the spectrum. In the experimental data, one expects a weak avoided level crossing to show up as a resonance in the inverse quality factor of the resonator.}
\label{fig:DampedDecoherentTransmission}
\end{figure}

\begin{figure}[htp]
\centering
\begin{minipage}{.23\textwidth}
    \centering
     \includegraphics[width=1\linewidth,trim={0.2cm 0.2cm 0.5cm 0.8cm},clip]{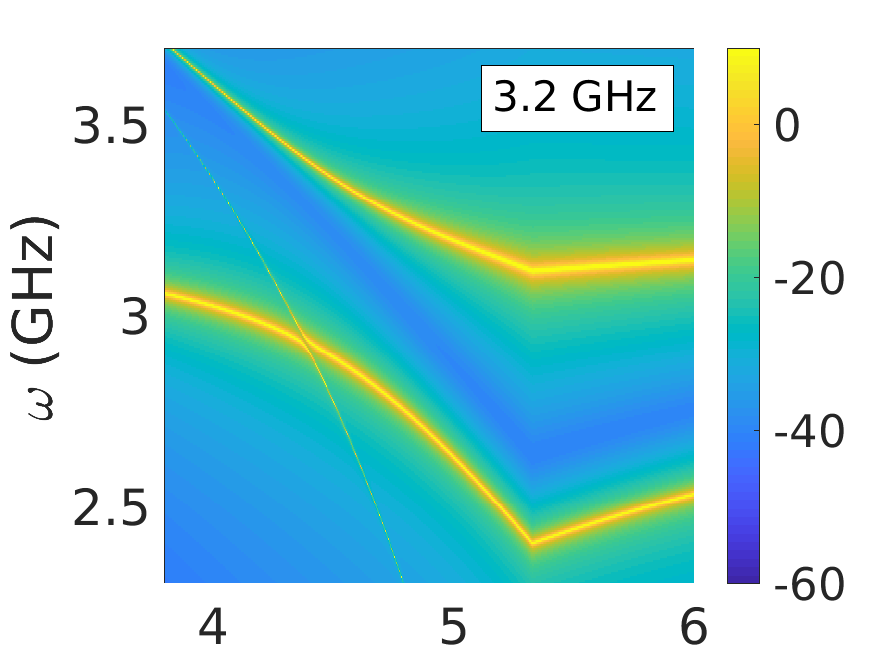} 
\end{minipage}
\begin{minipage}{.23\textwidth}
    \centering
     \includegraphics[width=1\linewidth,trim={0.1cm 0.2cm 0.3cm 0.8cm},clip]{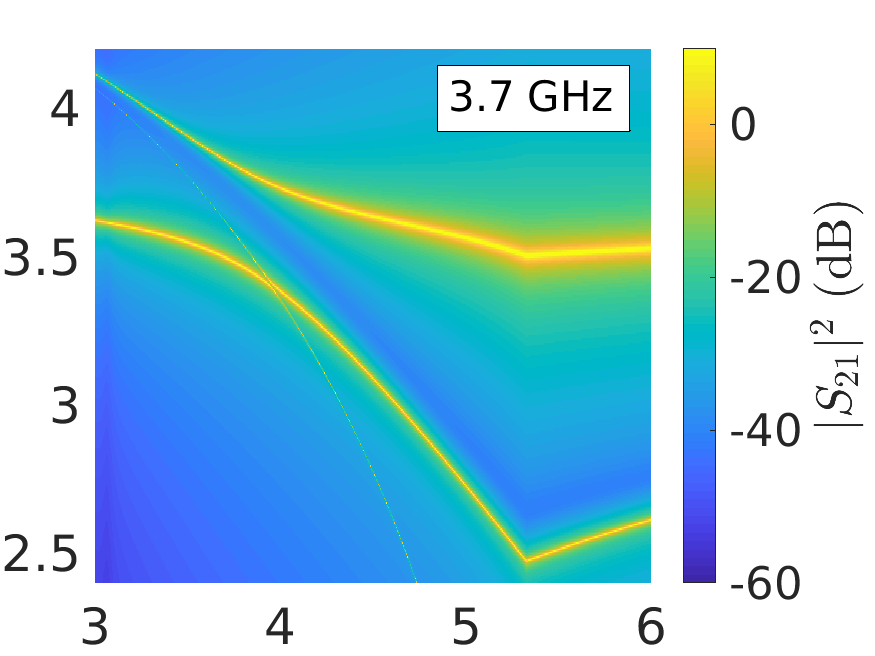} 
\end{minipage}
\begin{minipage}{.23\textwidth}
    \centering
     \includegraphics[width=1\linewidth,trim={0.2cm 0.2cm 0.5cm 0.5cm},clip]{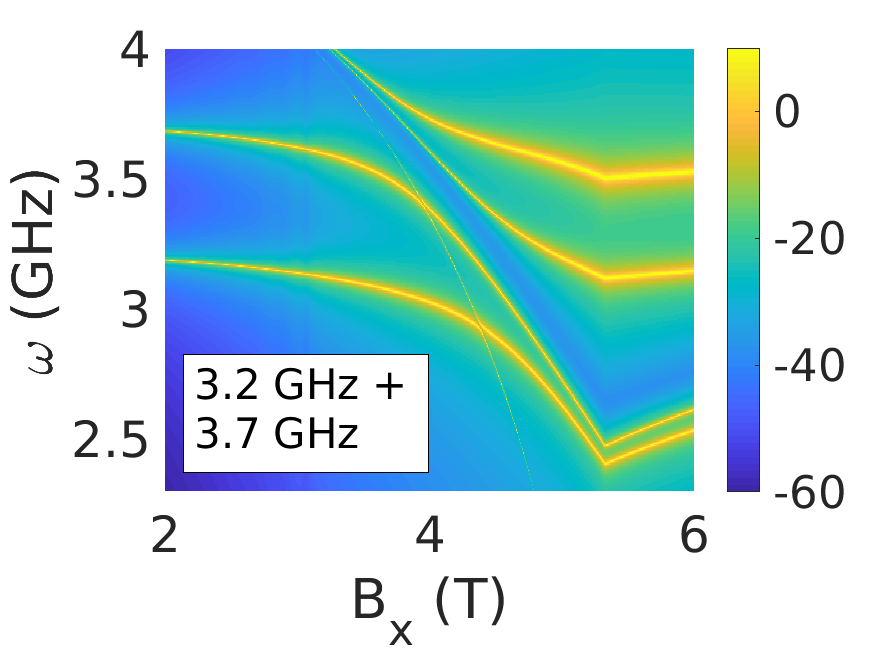} 
\end{minipage}
\begin{minipage}{.23\textwidth}
    \centering
     \includegraphics[width=1\linewidth,trim={0.1cm 0.2cm 0.3cm 0.5cm},clip]{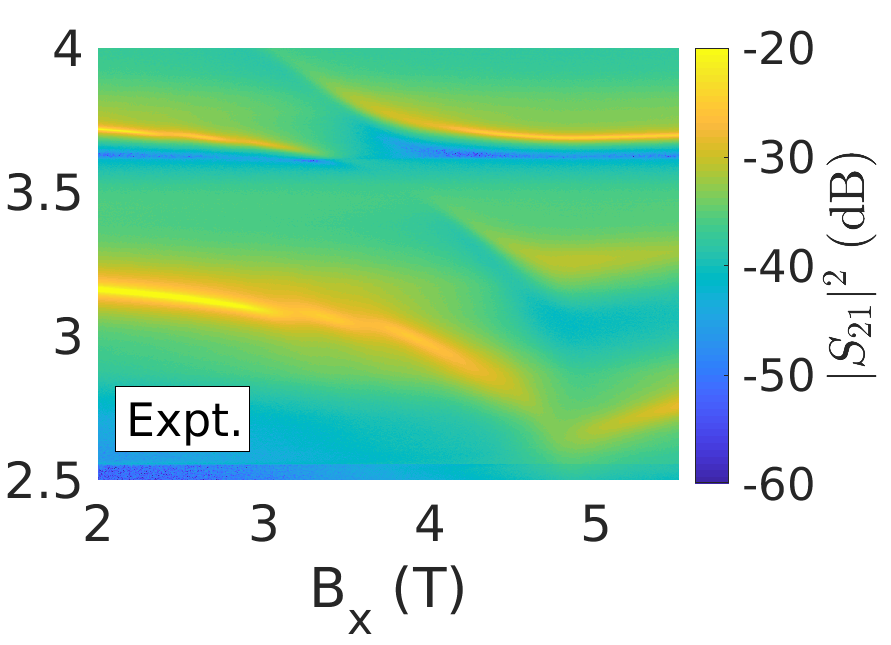} 
\end{minipage}
\caption{Mixed single ion and collective mode transmission of LiHoF$_4$ in a 3.2GHz and 3.7GHz resonator at zero temperature. The filling factor is set to $\eta=0.25$, and the damping parameters are chosen to be in line with what one expects for spin vacancies in diamond. The decoherence factor is set to $\gamma_{dec}=100GHz$, a value for which, although faint, the soft mode is visible in the transmission spectrum. Comparing the avoided level crossing in the $3.2GHz$ resonator to the $3.7GHz$ resonator in the upper pair of figures, we see a larger avoided level crossing at the lower frequency. This is due to the increase of the spectral weight of the magnon mode at $3.2GHz$, which supersedes the reduction in coupling strength due to the lower resonator frequency. In the lower pair of figures, we sum the calculated transmission from the $3.2GHz$ and the $3.7GHz$ resonators, and compare the results to transmission through a bimodal loop gap resonator. In the experimental data, interactions between the resonator modes lead to an antiresonance near $3.6GHz$ and hybridization of the polariton modes not accounted for in the theoretical calculation. The lowest polariton mode in the experimental data exhibits weak avoided level crossings consistent with the presence of the collective soft mode, and Walker modes, in the material (see text for details).}
\label{fig:DampedDecoherentTransmission2}
\end{figure}

When the coupling between the resonator photons and the magnetic excitations is weak, the resonator transmission spectrum does not exhibit any novel features. If the filling factor is increased to $\eta=0.25$, we see interesting features in the RPA transmission spectrum due to the hybridization of the magnon and photon modes. In Figures \ref{fig:DampedTransmission} - \ref{fig:DampedDecoherentTransmission2}, we show the effects of damping and decoherence on the theoretical resonator transmission, and we compare the results to experimental data. 

In Fig. \ref{fig:DampedTransmission}, we consider constant ohmic damping of the magnon-modes. As discussed in Section \ref{sec:DampedProp}, we find that strong damping of the magnon modes may prevent the superradiant quantum phase transition expected as the quantum Ising material is tuned through its critical point. The theoretical results for the resonator transmission are in poor agreement with the experimental data, shown in Fig. \ref{fig:ExptTransmission}, and it is necessary to refine our treatment of the damped magnon modes. In Fig. \ref{fig:DampedModes}, we show the single ion and collective mode resonator transmission using a more realistic model for the damping parameters. Our estimates of the magnitudes of the damping parameters fall short of what is necessary to prevent superradiance. In order to account for this discrepancy, we introduce a phenomenological model to account for decoherence of the collective magnon modes.

To explore the effects of decoherence, we assume spectral weight is transferred from collective magnon modes to single ion excitations in the magnon-polariton propagator. This leads to a reduced coupling between the collective magnon modes and the photons, and mixed single ion and collective mode transmission in the resonator. In Fig. \ref{fig:DampedDecoherentTransmission}, we show the effects of tuning the decoherence rate of the collective magnon modes in the phenomenological model; as the decoherence rate is increased, the superradiant quantum phase transition gives way to an avoided level crossing between the magnon soft mode and the resonator mode, which, upon further increasing the decoherence rate, gives way to a resonance at the transverse field values where the resonator mode is degenerate with the soft mode. The model is then used to calculate mixed single ion and collective mode resonator transmission at frequencies where the resonator mode is degenerate with the lowest  single ion excitation, and the results are compared to experimental data for a bimodal loop gap microwave resonator. We find good agreement between the experimental data and the theoretical results. 

Consider Fig. \ref{fig:DampedTransmission}, in which we show the zero temperature transmission spectrum of LiHoF$_4$ in a $\omega_r/(2\pi) = 1.9GHz$ resonator with constant ohmic damping of the magnon modes. We see that when the modes are weakly damped, the lower polariton mode softens to zero marking a superradiant quantum phase transition in the system. As discussed following equation (\ref{eq:gmDamped}), by increasing the damping of the soft mode from $\Gamma_{m=1} = 1 \mu K=20.837kHz$ to $\Gamma_{m=1} = 0.5K=10.419GHz$, the lower polariton mode no longer softens to zero; however, the resulting transmission function is in poor agreement with the experimental data. 

In Fig. \ref{fig:ExptTransmission} we show the experimental resonator transmission and the inverse quality factor of the resonator mode ($1/Q$), which is proportional to the linewidth of the polariton mode. The inverse quality factor shows a resonance near the phase transition that may be decomposed into the sum of three distinct peaks. The central peak is due to absorption at the phase transition. As the LiHoF$_4$ sample is tuned through its critical point, one expects absorption at all frequencies, similar to critical opalescence \cite{LiberskySM}. The two satellite peaks correspond to resonances in the transmission function where the resonator polariton mode ($\omega_p$) is degenerate with the soft mode ($\omega_m$). To better capture the experimental data, we consider a refined model for the damping parameters, and make use of an ansatz meant to capture the effects of decoherence of the collective magnon modes.

In a more realistic model for the damping of the magnon modes, the damping parameters will vary as a function of transverse field and frequency. We neglect the memory effects associated with the frequency dependence of the damping parameters; however, we incorporate the transverse field dependence of the parameters by considering damping due to an oscillator bath environment at the frequency of the magnon mode. The damping of a mode at frequency $\omega_m$ is given by $\Gamma_m = \gamma'(\omega_m)$, where $\gamma'(\omega)$ is given in equation (\ref{eq:gamma}) of Appendix \ref{ap:CHO}. We find that
\begin{align}
\Gamma_m = 2\pi \sum_z g_{zm}^2 [\delta(\omega_m-\omega_z)-\delta(\omega_m+\omega_z)], 
\end{align}
where the frequency independent damping function $\Gamma_m$ is in agreement with what one obtains using a master equation approach \cite{Clerk}; the transverse field dependence of the damping function is due to the transverse field dependence of the magnon mode $\omega_m$. Converting the sum over bath modes to an integral, one obtains
\begin{align}
\Gamma_m = 2\pi g_{zm}^2 \rho_b(\omega_m) n(\omega_m),
\end{align}
where $\rho_b(\omega_m)$ is the density of states of the bath modes at frequency $\omega_m=\omega_z>0$, and $n(\omega_m)$ is the Bose-Einstein distribution function. 

Recall that the magnon-photon coupling strength in LiHoF$_4$ is given by $g_m^2=\alpha^2 A_m$ (equation (\ref{eq:coupling})), with $\alpha^2 = 2\pi \eta^2 (\rho J_D)  \omega_r $, as in equation (\ref{eq:BareCoupling2}). We assume the coupling between magnons and bath modes has a similar form $g_{zm}^2 = g_0^2 A_m = Z \omega_m A_m$ (for light-matter coupling, one has $Z=2\pi \eta^2 \rho J_D$). Assuming a quadratic density of states, $\rho_b(\omega_m) = \rho_0 \omega_m^2$, in the high temperature limit ($\beta \omega_m \ll 1$), the damping parameter may be written
\begin{align}
\label{eq:Damping}
\Gamma_m = C_0 A_m \omega_m^2 \quad \text{where} \quad
C_0 = 2\pi Z \hbar \rho_0  k_B T.
\end{align}
The spectral weights of the magnon modes go like $A_m \sim 1/\omega_m$, so one expects a reduction in the damping of the soft mode as $\omega_m \rightarrow 0$.

The damping of LiHoF$_4$ due to a phonon bath has been analyzed by Buchhold \cite{Buchhold} \textit{et al.} Specific heat measurements \cite{Aggarwal} in LiYF$_4$ and LiLuF$_4$ indicate Debye temperatures of $\theta_D=560K$ and $\theta_D=540K$, respectively. The Debye temperature of LiHoF$_4$ is expected to be similar. In terms of the Debye temperature and the corresponding Debye frequency $\omega_D = k_B \theta_D /\hbar$, and assuming the phonon density of states is $\rho_{ph}(\omega_m) = \rho_0 \omega_m^2$, Buchhold \textit{et al}. find the damping of a magnon mode $\omega_m$ to be
\begin{align}
\Gamma_m \approx \gamma_D \frac{T}{\theta_D \omega_D^2} \omega_m^2 
= \widetilde{\gamma}_D \frac{T}{\theta_D \omega_D^2} A_m \omega_m^2,
\end{align} 
where in the final expression we have excluded $A_m$ from the decay rate at the Debye frequency and temperature, $\gamma_D$. Comparing with equation (\ref{eq:Damping}), we see that $C_0 = \widetilde{\gamma}_D T/ (\theta_D \omega_D^2)$. Little information on phonons in LiHoF$_4$ is available; however, for spin vacancies in diamond one has \cite{Buchhold} $\gamma_D / (\theta_D \omega_D^2) = 10^{-6} \rightarrow 10^{-5}\ (GHz\ K)^{-1}$. At the experimentally relevant temperature of $T=50mK$, this leads to $C_0 \approx 5 \times (10^{-8}\rightarrow 10^{-7}) GHz^{-1}$, and damping of the magnon modes of less than a kilohertz. This is very weak damping of the modes; however, interactions between magnetic fluctuations, and environmental degrees of freedom other than phonons are expected to increase the dampings. 

In Fig. \ref{fig:DampedModes}, we consider single ion and collective mode resonator transmission, with the damping of the collective modes given in (\ref{eq:Damping}), and the damping of the single ion excitations given by $\Gamma_{mn} = C_0 a_{mn} E_{nm}^2$, where $a_{mn}$ and $E_{nm}$ are discussed in Section \ref{sec:MF}. We consider a $3.2GHz$ resonator relevant to the experimental data shown in Fig. \ref{fig:DampedDecoherentTransmission2}; results for a $1.9GHz$ resonator are similar. The single ion resonator transmission follows from replacing $\chi$ with $\chi_0$ in equations (\ref{eq:wmp}) and (\ref{eq:Gammamp}), as discussed in Section \ref{sec:MF}. We set $C_0=10^{-5}\ K^{-1} = 4.8 \times 10^{-7}\ GHz^{-1}$, so the damping parameters are roughly in line with what one expects for spin vacancies in diamond. The MF and RPA resonator transmission is calculated at zero temperature, which accurately captures the most dominant modes present at the experimentally relevant temperature of $T=50mK$. This validates using the $T=50mK$ estimate for the damping parameters in the zero temperature resonator transmission calculations. Modes corresponding to excitations between thermally excited states of the quantum Ising material will be the subject of future work.

The damping of the collective magnon modes in Fig. \ref{fig:DampedModes} is insufficient to prevent a superradiant phase transition in the system, which is inconsistent with the experimental data. However, we have not accounted for the quantum coherence of the collective magnon modes. We attempt to do so by assuming that spectral weight is transferred from the collective magnon modes to the single ion excitation spectrum shown on the right hand side of Fig. \ref{fig:DampedModes}. Indeed, for each mode in equations (\ref{eq:wmp}) and (\ref{eq:Gammamp}), we assume (for example)
\begin{align}
& \qquad\qquad \frac{2g_m^2 \omega_r (\omega-\omega_m)}{(\omega-\omega_m)^2+(\Gamma_m/2)^2} \rightarrow
\\ \nonumber
&\frac{2 \widetilde{g}_m^2 \omega_r (\omega-\omega_m)}{(\omega-\omega_m)^2+(\Gamma_m/2)^2}
+ \frac{2 \widetilde{g}_{mn}^2 \omega_r (\omega-E_{nm})}{(\omega-E_{nm})^2+(\Gamma_{mn}/2)^2},
\end{align}
where
\begin{align}
\widetilde{g}_m^2(\omega=\omega_m) 
= \alpha^2 A_m \biggr[\frac{\omega_m^2}{\gamma_{dec}^2+\omega_m^2}\biggr]
\end{align}
and
\begin{align}
\widetilde{g}_{mn}^2(\omega=\omega_m) = \alpha^2 a_{mn} \biggr[1-\frac{\omega_m^2}{\gamma_{dec}^2+\omega_m^2}\biggr].
\end{align}
This leads to mixed single ion and collective mode transmission in the magnon-polariton propagator. Fourier transforming $\widetilde{g}_m^2(\omega)$, one finds that this ansatz corresponds to exponential decay of the collective mode spectral weight at a rate determined by $\gamma_{dec}$. We set $\omega=\omega_m$ to capture decoherence at the relevant frequency scale of the quantum Ising material. In the development of the magnon-polariton theory, the photons couple to an auxiliary field which describes magnetic fluctuations, and determines the magnon modes present in the material. The quantum coherence of the collective magnon modes is a tacit assumption which may not be valid if environmental degrees of freedom, or higher order interactions (beyond the RPA) between the magnetic fluctuations, lead to decoherence on timescales faster than the relevant timescales of the magnon modes.

In Fig. \ref{fig:DampedDecoherentTransmission}, we show the mixed single ion and collective mode resonator transmission as one tunes the decoherence rate $\gamma_{dec}$. The damping parameters are chosen to be roughly consistent with what one expects for spin vacancies in diamond, as in Fig. \ref{fig:DampedModes}. With $\gamma_{dec} = 0.5GHz$, the reduction in the coupling strength is insufficient to prevent the superradiant quantum phase transition. Increasing the decoherence rate to $\gamma_{dec}=15GHz$, which is larger than the relevant magnon and resonator mode frequency, leads to an avoided level crossing in the transmission spectrum, rather than a superradiant phase transition, as shown in the upper right plot in Fig. \ref{fig:DampedDecoherentTransmission}. Further increasing the decoherence rate attenuates the soft mode, and weakens the avoided level crossing in the transmission spectrum. With the coherence time set to picoseconds, which is shorter than the timescale set by the inverse of the magnon mode frequency, the magnon soft mode will show up as a resonance in the resonator transmission spectrum, as seen in the experimental data. 

We have shown the effects of ohmic damping of the magnon modes, in conjunction with an ansatz meant to capture the impact of decoherence of the collective spin excitations comprising the magnon modes. When decoherence is accounted for, we find that the superradiant quantum phase transition, or strong avoided level crossing, expected as the spectral weight of the magnon soft mode diverges, gives way to a resonance in the resonator transmission function. In Fig. \ref{fig:DampedDecoherentTransmission2}, we consider resonator frequencies of $\omega_r/(2\pi) = 3.2GHz$ and $3.7GHz$, and compare the calculated transmission function to experimental data for a bimodal loop gap resonator. At these frequencies, the resonator modes are degenerate with the lowest single ion excitation in the system. We see strong avoided level crossings when the lowest single ion excitation is degenerate with the resonator modes. The increased spectral weight of the single ion excitation at 3.2GHz leads to a stronger avoided level crossing than at 3.7GHZ, despite the reduction in frequency. This is consistent with the avoided level crossings seen in the experimental data.

The experimental data is for a bimodal resonator. In the theoretical calculation we assume the two resonator modes are independent, and sum their response. This fails to capture interactions between the resonator modes, which lead to the antiresonance seen in the experimental data near $3.6GHz$, and mixing of the calculated polariton modes. Nevertheless, we find good agreement between the calculated resonator transmission and the experimental data. The lowest polariton mode exhibits a series of weak avoided level crossings in the ferromagnetic phase of the quantum Ising material. These avoided level crossings are due to the soft mode, and Walker modes present in the material \cite{Walker1, Walker2}. An analysis of the Walker modes will be the subject of future work. Previously, in reference \cite{LiberskySM}, the magnon mode responsible for the avoided level crossings seen in the experimental data shown in Fig. \ref{fig:DampedDecoherentTransmission2} was attributed to an excited state transition; this was based on an RPA analysis of the LiHoF$_4$ crystal. In the current work, assuming mixed single ion and collective mode resonator transmission, we attribute these avoided level crossings to the lowest single ion excitation (ground state to first excited state) shown by the dashed line in the inset to Fig. \ref{fig:ModesWeakCoupling}. The structure and energy of the lowest single ion excitation, and the first excited state in the RPA calculation, are similar.

Accounting for, and exploring, the effects of dissipation when polariton modes are coupled to environmental degrees of freedom is an active research area \cite{WangHu, CorteseDeLiberato}; incorporating the effects of decoherence of the collective magnon modes comprising the magnon-polaritons in a quantum Ising material coupled to a resonator mode is a more difficult task. Here, we have developed a basic formalism amenable to investigating these problems in real materials, with LiHoF$_4$ being the magnetic system of primary interest. We have demonstrated the effects of ohmic damping of the magnon modes in LiHoF$_4$ in a microwave resonator, and we have explored the consequences of decoherence of the magnon modes present in the material via an ansatz in which spectral weight is transferred from the collective modes to single ion excitations. Our results are in good agreement with experimental data for LiHoF$_4$ in a microwave resonator; we leave further refinements of the theory, and more sophisticated numerical analysis, as a subject for future work.

\section{Conclusions and Outlook}

Beginning with a microscopic spin model for a quantum Ising system in a microwave resonator, we have derived an effective finite temperature quantum field theory for the magnon-photon system, and an effective Hamiltonian for the coupled bosonic modes. The theory has been used to calculate the magnon-polariton propagator, and the results have been applied to LiHoF$_4$, which has a complex, multilevel, single site Hamiltonian. One may also apply this formalism to the quantum optics models, and quantum environment models, discussed in Appendix \ref{ap:Models}. 

Our analysis of a quantum Ising material via the introduction of an auxiliary field describing the magnetic fluctuations goes beyond standard spin quantization techniques. The resulting theory captures multiple magnon modes, the quasi-elastic diffusive pole of the quantum Ising material, and excitations between thermally excited states of the material. Our treatment of the light in terms of harmonic oscillator variables is basic; however, we believe it provides clarity, and we have made contact between paradigmatic quantum optics models and oscillator bath theory. One may extend and expound details of the theory by treating the light, or the environment, in a more sophisticated manner.

A key result of this paper is that tuning the applied transverse field allows one to tune the magnon-photon coupling strength. As one approaches the critical point of the quantum Ising material, the magnon-photon coupling strength will diverge. A fixed system of spins in an ac magnetic field  will not exhibit a diamagnetic response, so deep strong coupling between the magnons and photons is achieved without the light-matter decoupling inherent in the Dicke \cite{Dicke}, Dicke-Ising \cite{Cortese}, and Hopfield models \cite{Hopfield}. However, in the real-world, coupling to an environment will lead to dissipation and decoherence, which may lead to weak coupling between the magnon and photon modes. We have treated dissipation and decoherence phenomenologically, and compared the results of the theory to experimental data on LiHoF$_4$ in a loop gap microwave resonator. We consider the agreement between the experimental data and the theoretical results to be good, although further refinement of the theory, particularly more detailed modeling of the environment and the resulting decoherence, and a more sophisticated numerical analysis, would be beneficial. We leave this a subject for future work.

We have focused on the magnon-polariton propagator because it may be the best way to make contact between theoretical work and experimental results, and our treatment of the light in terms of harmonic oscillator variables is the easiest way to obtain results. Harmonic oscillator variables, and eigenstates, have also been used to study entanglement, and the quantum-chaotic properties, of the Dicke model \cite{EmaryBrandesPRE, Lambert}. Alternatively, one may make use of a coherent state basis of eigenstates for the light. This was an original approach to the problem \cite{Wang} that allows for a more thorough investigation of the thermodynamics of the system. More recently, coherent states were used to study entanglement between a qubit and a field mode \cite{Everitt}. We see further investigation into the entanglement properties of light-matter systems as a promising area for research, with particular relevance to high sensitivity magnon detection \cite{LQScience}, and associated quantum technologies.

This work provides a detailed microscopic theory of a quantum optics system. Such a theory is necessary in order to make progress in more topical research areas such as the non-equilibrium phases and phase transitions \cite{Sieberer, DallaTorre, Kirton}, and novel dynamics \cite{Henriet, Hanai}, present in damped-driven quantum systems \cite{Kasprzak, Muniz}. The formalism here is complementary to, and more general than, standard approaches which make use of bosonic or fermionic representations of the spin degrees of freedom, and the field theory is amenable to treatment via the Keldysh functional integral approach. Finally, the burgeoning field of quantum magnonics will require models of light-matter interactions, following the lines of the present investigation.

\section{Acknowledgments}

The authors would like to thank Yikai Yang and Philip Stamp for helpful discussions. Experimental work at Caltech was supported by U.S. Department of Energy  Basic Energy Sciences, Award No. DE-SC0014866. 

\begin{appendices}

\section{Other Models}
\label{ap:Models}

In the absence of spin-spin interactions, our model shares similarities with the Dicke model \cite{Dicke}, which is a paradigmatic model of quantum optics. The basic Dicke model describes an atomic cloud, approximated as a set of two level systems, coupled to a single photonic field mode ($\hbar=1$) 
\begin{align}
\mathcal{H}_{Dicke} = \omega_r a^{\dagger} a + \omega_0 J^z + \frac{\alpha}{\sqrt{N}} (a^{\dagger}+a) J^x
+ \mathcal{H}_{A^2}.
\end{align}
The collective atomic operators are given by $J^{\mu} = \sum_i J_i^{\mu}= \sum_i \sigma_i^{\mu}/2$, where the $\sigma_i^{\mu}$ are Pauli operators. In this model, the atoms (or spins) are mobile charged particles. This collective set of atomic operators couples to a position operator of a single field mode, $x \sim a^{\dagger}+a$. The diamagnetic term, 
\begin{align}
\label{eq:A2}
\mathcal{H}_{A^2} = D (a^{\dagger}+a)^2, 
\end{align}
comes from squaring the canonical momentum of the mobile charged particles. Invoking the Thomas-Reiche-Kuhn sum rule for a multilevel atom \cite{Rzazewski, Nataf}, one finds that in the two level approximation $D > \alpha^2/\omega_0$, so that the magnitude of the diamagnetic term diverges like the square of the coupling strength. If Ising interactions between the atoms in the Dicke model are included, one has the Dicke-Ising model \cite{Cortese}. The absence of the diamagnetic term in equation (\ref{eq:Hx}) distinguishes it from the Dicke-Ising model. As the diamagnetic term has important consequences, these two models should be considered distinct.

The Dicke model was introduced in 1954 to describe an atomic system in a light field. In 1958, Hopfield developed a model for dielectric materials in an electromagnetic field \cite{Hopfield}. Considering only a pair of modes in the resulting exciton-polariton theory, the Hopfield model is given by ($\hbar=1$)
\begin{align}
\label{eq:Hop}
\mathcal{H}_{Hop.} = \omega_r a^{\dagger} a + \omega_0 b^{\dagger} b - ig (a^{\dagger}+a) (b^{\dagger}-b)
+ \mathcal{H}_{A^2}.
\end{align}
The coupling in the Hopfield model is between an effective position operator, $x\sim a^{\dagger}+a$, and an effective momentum operator $p \sim i(b^{\dagger}-b)$. The diamagnetic term is the same as for the Dicke model (equation \ref{eq:A2}), with $D = g^2/\omega_0$. As for the spin-photon model in equation (\ref{eq:Hp}), one can show that the substitution $i(b^{\dagger}-b) \rightarrow b^{\dagger}+b$ leads to an equivalent formulation of the model (see Appendix \ref{ap:coupling}).  The formalism developed here encompasses both the Dicke and Hopfield models, and generalizes the basic Dicke model to include interactions between multilevel spins or atoms.

The models developed by Dicke and Hopfield share a connection with work on quantum environments. In quantum optics, light is an intrinsic part of the system; in the theory of quantum environments, light, other bosonic or fermionic modes, and spin degrees of freedom, are extrinsic to the system of interest, and lead to dissipation and decoherence in the system. The quantum optics models discussed above share strong similarities with standard decoherence models describing a quantum system coupled to its environment, such as the Caldeira-Leggett model and the spin-boson model \cite{CaldeiraLeggett, LeggettSB}. In the decoherence models the system is comprised by the matter modes, and the environment is analogous to the light in the quantum optics models.

The formalism developed in this work is applicable to all the models discussed above. Furthermore, it can be used to generalize the basic Dicke model and spin-boson model to include interactions between multilevel atoms or spins with complicated single ion Hamiltonians. 

\section{Momentum verses Position Coupling}
\label{ap:coupling}

The magnon-polariton theory has been derived for a system in which the spins couple to a photon position operator, as in equation (\ref{eq:Hx}). One can show that equation (\ref{eq:Hp}) is an equivalent formulation of the model. The two expressions are related by a canonical transformation that swaps photon position and momentum operators \cite{Leggett84}. Similarly, the position-momentum coupling in the Hopfield model, given by equation (\ref{eq:Hop}), may be replaced with a coupling between position operators.

Consider the Hopfield model. In terms of harmonic oscillator variables,
\begin{align}
x = \sqrt{\frac{\hbar}{2m\omega}} (a^{\dagger}+a) \quad \text{and} \quad
p = i \sqrt{\frac{\hbar m \omega}{2}} (a^{\dagger}-a),
\end{align}
the model is written
\begin{align}
\mathcal{H}_{Hop.} =\frac{P^2}{2M} + \frac{1}{2}M\omega_r^2 X^2 
+ \frac{(p-cmX)^2}{2m} + \frac{1}{2}m\omega_0^2 x^2,
\end{align}
with $c = 2g \sqrt{M\omega_r/m\omega_0}$. One may replace the position-momentum coupling with a position coupling by making the change of variables $\widetilde{x} = p/(m\omega_0)$ and $\widetilde{p} = -m\omega_0 x$. In terms of creation and annihilation operators, this canonical transformation leads to
\begin{align}
\widetilde{\mathcal{H}}_{Hop.} = \hbar \omega_r &\biggr(a^{\dagger} a + \frac{1}{2}\biggr) 
+ \hbar \omega_0 \biggr(\widetilde{b}^{\dagger} \widetilde{b} + \frac{1}{2}\biggr)
\\ \nonumber
&- \hbar g (a^{\dagger}+a)(\widetilde{b}^{\dagger}+\widetilde{b}) 
+ \frac{\hbar g^2}{\omega_0}(a^{\dagger}+a)^2,
\end{align}
where the coupling is now between position operators. Note that this Hamiltonian is equivalent to the Caldeira-Leggett Hamiltonian (see Appendix \ref{ap:CHO}) if only a single bath mode is considered. For reference, we note that the roles of the $a$ and $b$ bosons in the interaction and in the diamagnetic term may be interchanged making use of a gauge transformation \cite{Garziano}.

Now consider the model given by equation (\ref{eq:Hp}). One may develop the magnon-polariton theory in the same manner as for the position coupling case. The photon component of the magnon-polariton Hamiltonian, $\mathcal{H}_{mp} = \mathcal{H}_{\gamma} + \mathcal{H}_{\phi} + \mathcal{H}_{int}$, is then
\begin{align}
\mathcal{H}_{\gamma} = \hbar \omega_r \biggr(a^{\dagger} a + \frac{1}{2}\biggr)
- i \hbar \lambda (a^{\dagger}-a) - D (a^{\dagger}-a)^2,
\end{align}
or, in terms of harmonic oscillator variables,
\begin{align}
\mathcal{H}_{\gamma} = \frac{p^2}{2m_r} + \frac{1}{2} m_r \omega_r^2 x^2
-\sqrt{\frac{2\hbar}{m_r \omega_r}} \lambda p + \frac{2}{m_r \hbar \omega_r} D p^2.
\end{align}
One may rescale the mass and frequency of the oscillators,
\begin{align}
m_{\gamma} = m_r \biggr[1+\frac{4D}{\hbar \omega_r}\biggr]^{-1} \quad \text{and} \quad
\omega_{\gamma} = \omega_r \sqrt{1+\frac{4D}{\hbar \omega_r}},
\end{align}
to obtain
\begin{align}
\mathcal{H}_{\gamma} = \frac{p^2}{2m_{\gamma}} + \frac{1}{2} m_{\gamma} \omega_{\gamma}^2 x^2
-\sqrt{\frac{2\hbar}{m_{\gamma} \omega_{\gamma}}} \lambda_{\gamma} p,
\end{align}
where $\lambda_{\gamma}$ is given in equation (\ref{eq:freqshift}).

When the spins couple to a photon momentum operator the interaction between the auxiliary field and the photons is given by
\begin{align}
\mathcal{H}_{int} = \hbar \alpha_{\gamma} \sqrt{\frac{2}{\hbar m_{\gamma} \omega_{\gamma}}} \phi_0 p,
\end{align}
with $\alpha_{\gamma}$ given in equation (\ref{eq:alpha}). Combining the terms involving photon operators, $\mathcal{H}_{\gamma \phi}=\mathcal{H}_{\gamma}+\mathcal{H}_{int}$, we have
\begin{align}
\mathcal{H}_{\gamma \phi} =  \frac{p^2}{2m_{\gamma}} 
+ \frac{1}{2} m_{\gamma} \omega_{\gamma}^2 x^2 - \frac{2\hbar}{m_{\gamma} \omega_{\gamma}} 
p (\lambda_{\gamma}-\alpha_{\gamma} \phi_0).
\end{align}
The canonical transformation between the photon position and momentum operators leads to
\begin{align}
\widetilde{\mathcal{H}}_{\gamma \phi} = \frac{\widetilde{p}^2}{2m_{\gamma}} 
+ \frac{1}{2} m_{\gamma} \omega_{\gamma}^2 \widetilde{x}^2 
- \sqrt{2\hbar m_{\gamma} \omega_{\gamma}} \widetilde{x} 
(\lambda_{\gamma} - \alpha_{\gamma} \phi_0), 
\end{align}
which is equivalent to the result obtained if the spins are coupled to photon position operators in the original Hamiltonian (the rescaled mass of the oscillator does not affect the quantized theory).

\section{The L$\text{i}$H$\text{o}$F$_4$ System}
\label{ap:LiHo}

Consider the low temperature effective Hamiltonian of LiHoF$_4$ given in equation (\ref{eq:LiHo}) of the main text. The truncation of the LiHoF$_4$ system, and the low energy electronuclear modes present in the system, have been dealt with in detail elsewhere \cite{Chakraborty, Tabei, MckenzieStamp, EisenlohrVojta}; here we present details relevant to the calculation of the magnon-polariton propagator.

In the random phase approximation (RPA), the longitudinal dynamic susceptibility may be written as $\chi(z) = \chi_0(z)/(1-V_0 \chi_0(z))$, where the mean field (MF) susceptibility, $\chi_0(z) = \widetilde{\chi}_0(z) + \chi_{el}^0 \delta_{z,0}$, is written explicitly in terms of the MF parameters of the system in equation (\ref{eq:chi0}). The inelastic component of the RPA susceptibility is $\widetilde{\chi}(z) = \widetilde{\chi}_0(z)/(1-V_0 \widetilde{\chi}_0(z))$, and the RPA expression for the quasi-elastic diffusive pole is
\begin{align}
\chi_{el} = \frac{\widetilde{\chi}_0(0)+\chi_{el}^0}
{1-V_0 (\widetilde{\chi}_0(0)+\chi_{el}^0)} - \frac{\widetilde{\chi}_0(0)}{1-V_0 \widetilde{\chi}_0(0)}.
\end{align}
Defining the ratio of the MF and RPA modes of the system to be
\begin{align}
R \equiv \frac{1}{1-V_0 \widetilde{\chi}_0(0)} = \frac{\prod_{n>m} E_{nm}^2}{\prod_m \omega_m^2},
\end{align}
the elastic component of the RPA susceptibility may be written
\begin{align}
\chi_{el} = \frac{R^2 \chi_{el}^0}{1-R V_0 \chi_{el}^0}.
\end{align}
The elastic component of the dynamic susceptibility has not been analyzed explicitly in this work; however, it is provided here for reference.

The inelastic component of the dynamic susceptibility, given in equation (\ref{eq:specrep}), determines the RPA modes of the LiHoF$_4$ system and their spectral weights. These spectral weights determine the strength of the magnon-photon coupling in the magnon-polariton theory. In terms of the MF energy levels and matrix elements of the longitudinal spin operator, the RPA expression for the inelastic component of the longitudinal dynamic susceptibility at zero wavevector is
\begin{widetext}
\begin{align}
\widetilde{\chi}(z) 
= \frac{-C_{zz}^2\sum_{n>m}|c_{mn}|^2 p_{mn} 2E_{nm} \prod_{t>s \neq nm} (E_{ts}^2-z^2)}
{\prod_{n>m} (E_{nm}^2-z^2)-C_{zz}^2 V_0\sum_{n>m}|c_{mn}|^2 p_{mn} 2E_{nm} 
\prod_{ts \neq mn} (E_{ts}^2-z^2)}.
\end{align}
\end{widetext}
In a needle shaped sample of LiHoF$_4$, with zero demagnetization field, the zero wavevector component of the interaction strength (equation (\ref{eq:V})), is approximately $V_0 \approx 74mK$, and, as mentioned following equation (\ref{eq:V}), $C_{zz}$ is a truncation parameter, with $J^z=C_{zz}\tau^z$ in the truncated spin-1/2 electronic subspace. The remaining parameters are defined in Section \ref{sec:DS} following equation (\ref{eq:chi0}).

The poles of the dynamic susceptibility determine the RPA magnon modes of the system, and their residues determine the spectral weights of the modes. The poles and residues can be calculated as in Section \ref{sec:RPA} of the body of the paper. One finds the spectral weights of the RPA magnon modes to be given by
\begin{align}
A_m = \frac{C_{zz}^2}{\omega_m}\sum_{n>m} |c_{mn}|^2 p_{mn} E_{nm}
\frac{\prod_{t>s \neq nm} [E_{ts}^2-\omega_m^2]}
{\prod_{s \neq m} [\omega_s^2-\omega_m^2]},
\end{align}
where $\{\omega_m \}$ are the zero wavevector RPA modes of the system. The spectral weights of the RPA modes scale like $A_m \sim 1/\omega_m$, with the spectral weight of the soft mode diverging at the quantum critical point.

\section{Coupled Harmonic Oscillators}
\label{ap:CHO}

Consider the Caldeira-Leggett Hamiltonian in which a harmonic oscillator is linearly coupled to a bath of harmonic oscillators (quantum Brownian motion) \cite{CaldeiraLeggett, Weiss, BreuerBook}
\begin{align}
\mathcal{H}_{CL} = \frac{P^2}{2M} + &\frac{1}{2}M\omega_s^2 X^2 
+\sum_z \biggr[\frac{p_z^2}{2m} + \frac{1}{2}m\omega_z^2 x_z^2\biggr]
\\ \nonumber
& - \sum_z c_z X x_z + \sum_z \frac{c_z^2}{2m_z \omega_z^2} X^2.
\end{align}
The bath leads to damping and decoherence of the primary oscillator. In terms of bosonic creation and annihilation operators the Caldeira-Leggett Hamiltonian may be written
\begin{align}
\label{eq:CL}
\mathcal{H}_{CL} = &\hbar \omega_s \biggr(b_0^{\dagger} b_0 + \frac{1}{2}\biggr) 
+ \sum_z \hbar \omega_z \biggr(a_z^{\dagger} a_z + \frac{1}{2}\biggr)
\\ \nonumber
&- \sum_z \hbar g_z (a_z^{\dagger}+a_z) (b_0^{\dagger}+b_0) 
+ \sum_z D_z (b_0^{\dagger}+b_0)^2,
\end{align}
where
\begin{align}
g_z  = \frac{c_z}{2 \sqrt{m_z\omega_z M \omega_s}} \quad \text{and} \quad
D_z = \frac{\hbar g_z^2}{\omega_z}.
\end{align}
The Caldeira-Leggett counterterm is equivalent to the diamagnetic term present in light-matter Hamiltonians. In order to determine the damping due to the bath, we calculate the propagator of the primary oscillator.

The counterterm shifts the frequency of the primary harmonic oscillator
\begin{align}
\mathcal{H}_{CL} = \hbar \overline{\omega}_s \biggr(b^{\dagger} b &+ \frac{1}{2}\biggr) 
+ \sum_z \hbar \omega_z \biggr(a_z^{\dagger} a_z + \frac{1}{2}\biggr)
\\ \nonumber
&- \sum_z \hbar \overline{g}_z (a_z^{\dagger}+a_z) (b^{\dagger}+b),
\end{align}
where the rescaled coupling and shifted frequencies are
\begin{align}
\label{eq:renormalizedcoupling}
\overline{g}_z  = \frac{c_z}{2 \sqrt{m_z\omega_z M \overline{\omega}_s}}
\quad \text{and} \quad
\overline{\omega}_s^2  = \omega_s^2\biggr[1 + \frac{4D_z}{\hbar \omega_s} \biggr].
\end{align}
The propagator of the shifted oscillator modes is defined by
\begin{align}
D_b(\tau) = \bigr\langle T_{\tau} \bigr(b^{\dagger}(\tau)+b(\tau)\bigr) 
\bigr(b^{\dagger}+b\bigr) \bigr\rangle.
\end{align}
Treating the interaction between oscillators perturbatively using the Matsubara formalism, one finds the propagator of the primary oscillator in Matsubara frequency space to be
\begin{align}
D_b(i\omega_n) = -\frac{2\overline{\omega}_s}{\beta \hbar} 
\left[ \frac{1}{(i\omega_n)^2 - \overline{\omega}_s^2
- \sum_z \frac{4\overline{g}_z^2 \overline{\omega}_s \omega_z}{(i\omega_n)^2 - \omega_z^2}}\right].
\end{align}
This is equivalent to equation (\ref{eq:MPprop3}) of the main text apart from the fact that the counterterm has shifted the frequency of the primary oscillator.

Consider a single bath mode. The poles of the polariton propagator yield the upper and lower polariton modes
\begin{align}
\omega_{\pm}^2 = \frac{\overline{\omega}_s^2+\omega_z^2}{2} \pm 
\sqrt{\biggr(\frac{\overline{\omega}_s^2-\omega_z^2}{2}\biggr)^2
+4\overline{g}_z^2 \overline{\omega}_s\omega_z}.
\end{align}
In the absence of the counterterm ($D=0$), the lower polariton mode reaches zero at a critical value of
$g_z = \sqrt{\omega_s \omega_z}/2$. In a light-matter system, this coupling strength marks a superradiant quantum phase transition \cite{HeppLieb, Wang}. The presence of the counterterm forestalls this transition so that $\omega_- > 0$ for any value of $g_z$. The counterterm is also responsible for a decoupling of the light and matter modes (or system and bath modes) as the coupling strength is increased \cite{Rzazewski}. Indeed, consider what happens as the bare coupling between oscillators diverges, $c_z \rightarrow \infty$. The shifted frequency of the primary oscillator diverges linearly with the coupling, $\overline{\omega}_s \sim c_z$, and the rescaled coupling between the oscillators goes like $\overline{g}_z \sim c_z/\sqrt{\overline{w}_s} \sim \sqrt{c_z}$. Comparing the rescaled coupling strength to the shifted oscillator frequency we see that $\overline{\eta} \equiv \overline{g}_z/\overline{\omega}_s \sim 1/\sqrt{c_z} \rightarrow 0$. As the bare coupling between oscillators is increased, the bath mode will become an increasingly weak perturbation to the system.

We return now to the oscillator bath environment. In order to make contact with standard results, we express the propagator for the shifted system modes in terms of the propagator of the original modes of the system $D_b = \overline{\omega}_s D_{b_0} / \omega_s$. Analytically continuing to real frequencies ($D_{b_0}^{ret}(\omega) = \beta D_{b_0}(i\omega_n \rightarrow \omega + i0^+$), the retarded propagator of the original bosonic system modes may be written 
\begin{align}
D_{b_0}^{ret}(\omega) = -\frac{2\omega_s}{\hbar} 
\left[\frac{1}{\omega^2 +i \gamma \omega- \omega_s^2}\right],
\end{align}
where the damping function is
\begin{align}
\gamma(\omega) = \frac{i}{\omega} \biggr[ \sum_z\frac{4g_z^2\omega_s}{\omega_z}+ \lim_{\epsilon \rightarrow 0} 
\sum_z \frac{4g_z^2 \omega_s \omega_z}{\omega^2 + i\omega \epsilon - \omega_z^2} \biggr].
\end{align}
The real and imaginary parts of the damping function, $\gamma = \gamma' +i\gamma''$, are given by
\begin{align}
\omega \gamma''(\omega) = \sum_z\frac{4g_z^2\omega_s}{\omega_z}\biggr[ 
\frac{-\omega^2}{\omega_z^2-\omega^2}\biggr],
\end{align}
and
\begin{align}
\label{eq:gamma}
\omega \gamma' = 2\pi \omega_s \sum_z  g_z^2  
\biggr[\delta(\omega-\omega_z)-\delta(\omega+\omega_z)\biggr].
\end{align}
In terms of the original harmonic oscillator variables, the spectral density of the bath is defined by
\begin{align}
J(\omega) \equiv M \omega \gamma'(\omega) 
= \frac{\pi}{2} \sum_z \frac{c_z^2}{m_z \omega_z} \delta(\omega-\omega_z),
\end{align}
in agreement with the standard result. The counterterm (or equivalently, the diamagnetic term) eliminates a zero frequency shift in $\gamma''$. This term is absent in the magnon-polariton theory. In the magnon-polariton theory, the photons are considered to be the system, and the magnons, which are themselves subject to dissipation and decoherence, comprise a bath.

\end{appendices}


\bibliography{biblioMagnonics}

\end{document}